\documentclass[letterpaper, 10 pt, conference]{ieeeconf}
\usepackage{graphicx} 
\usepackage{multicol}
\usepackage{multirow}
\usepackage{blindtext}
\usepackage{amsmath, amssymb, xparse}
\usepackage{algorithm}
\usepackage{algpseudocode}
\usepackage{hyperref}
\usepackage{cleveref}
\usepackage{subcaption}
\usepackage[table,x11names]{xcolor}
\usepackage{longtable}
\usepackage{colortbl}
\definecolor{Gray}{gray}{0.8}
\definecolor{LightRed}{rgb}{1,0.85,0.85}
\definecolor{LightYellow}{rgb}{1,1,0.75}
\definecolor{LightGreen}{rgb}{0.8,0.97,0.8}

\title{\LARGE \bf
A switching Kalman filter approach to online mitigation and correction of sensor corruption for inertial navigation}
\author{Artem Mustaev, Nicholas Galioto, Matt Boler, John D. Jakeman, Cosmin Safta, Alex Gorodetsky}
\date{March 2024}

\newcommand{\numy}{n}
\newcommand{\nomobs}{\bar{H}}
\newcommand{\corobs}{\tilde{H}}

\newcommand{\yk}{\mathcal{Y}_k}

\newcommand{\K}{\mathcal{K}}

\newcommand{\rmd}{{\rm d}}
\newcommand{\reals}{\mathbb{R}}

\newcommand{\dimx}{d_x}
\newcommand{\dimy}{d_y}
\newcommand{\dimth}{d_{\theta}}

\newcommand{\vf}{\mathbf{f}}

\newcommand{\vg}{\mathbf{g}}
\newcommand{\vb}{\mathbf{b}}

\newcommand{\vv}{\mathbf{v}}
\newcommand{\vn}{\mathbf{n}}
\newcommand{\vX}{\mathbf{X}}

\newcommand{\ar}{\boldsymbol{\omega}^{b}}
\newcommand{\art}{\boldsymbol{\tilde{\omega}}^{b}}
\newcommand{\spf}{\mathbf{f}^{b}}
\newcommand{\spft}{\mathbf{\tilde{f}}^{b}}

\begin{document}

\maketitle
\thispagestyle{plain}
\pagestyle{plain}
\begin{abstract}
    This paper introduces a novel approach to detect and address faulty or corrupted external sensors in the context of inertial navigation by leveraging a switching Kalman Filter combined with parameter augmentation. Instead of discarding the corrupted data, the proposed method retains and processes it, running multiple observation models simultaneously and evaluating their likelihoods to accurately identify the true state of the system. We demonstrate the effectiveness of this approach to both identify the moment that a sensor becomes faulty and to correct for the resulting sensor behavior to maintain accurate estimates. We demonstrate our approach on an application of balloon navigation in the atmosphere and shuttle reentry. The results show that our method can accurately recover the true system state even in the presence of significant sensor bias, thereby improving the robustness and reliability of state estimation systems under challenging conditions. We also provide a statistical analysis of problem settings to determine when and where our method is most accurate and where it fails.
\end{abstract}

\section{Introduction}

\subsection{Motivation}
External corrections to inertial navigation systems (INS) are essential to realize highly performant vehicle navigation systems across a variety of domains~\cite{groves2015principles}. However, without  external sensor readings, the drift of Inertial Measurement Units (IMU) cause an accumulation of errors in estimates of the system state. External measurements such as those provided by global positioning systems (GPS) can be used to correct the drift inherent in IMU data.   However, GPS and other sensors operating external to the vehicle are not immune to failure or corruption. For example, external threats such as spoofing attacks can introduce biases that lead to inaccurate state estimates having severe consequences, for example altering the vehicle course or even fully take over the control of a vehicle~\cite{kerns2014unmanned,bhatti2017hostile,jafarnia2012gps}. Consequently, as the reliance critical systems on GPS grows, we must develop robust methodologies to detect, mitigate, and recover from GPS corruption and failure.

\subsection{Existing Approaches}
A variety of approaches exist for performing state estimation under the possibility of sensor failure, and these approaches include both preventive and reactive methods. Preventive measures use secure communication protocols and other protective strategies to limit the instances of GPS corruption \cite{haider2016survey,psiaki2016gnss}. However, these methods do not eliminate the risk entirely, and reactive mitigation strategies are necessary. While they are important, we do not focus on preventative approaches in this paper. In contrast, typical reactive approaches involve detecting the presence of corruption and then discarding the corrupted GPS measurements altogether~\cite{liang2022anti}. While this approach prevents misleading information from affecting the state estimation, it also results in the loss of potentially valuable information that could still contribute to accurate state estimation if appropriately managed. This approach, although common, limits the robustness of state estimation systems, especially in environments where GPS signals are intermittent or degraded.

Recent advancements in reactive methods have focused on developing more sophisticated methodologies that can continue to use GPS measurements when they are corrupted. Notably, particle filters have been employed to estimate the posterior distribution of the state by using a set of weighted particles, each representing a possible state of the system~\cite{majidi2018new}. This approach employs a hypothesis based method to verify whether the difference between GPS measurement and prediction of the IMU measurement is greater than certain threshold set based on the GPS accuracy. The algorithm proceeds with normal particle filtering for estimating the state until data corruption is detected. The particle weights are then dynamically adjusted or a model describing the assumed corruption process is added, thereby decreasing the influence of corrupted
measurements.

To mitigate the downsides of weighted filters, alternative approaches use model-based learning to learn the parametrized corruption bias model \cite{fan2017synchrophasor} to mitigate spoofing. For example, the above mentioned reference compares the likelihood of receiving the current measurement based on previous states to a set threshold to identify if spoofing is present. Once corruption is detected, the system assigns a cost function to the current measurement based on the model parameters. It then uses iterative techniques, such as the golden search method \cite{chong2013introduction}, to find a set of parameters that minimize the cost function and, therefore, are most likely to be the actual spoofing parameters. Once the parameters of the bias model are retrieved, the corruption is removed from the measurements, and state estimation proceeds as normal. This approach is significantly superior to the ones discussed previously as it identifies the corruption and recovers the initial measurements of the GPS, hence making use of all the available information for state estimation.  However, it can be expensive to implement and extend to other corruption models beyond specific forms considered in \cite{fan2017synchrophasor}.

\subsection{Paper Contributions}
In this paper, we introduce a novel approach that combines the Switching Kalman Filter (SKF) with parameter augmentation techniques to improve state estimation in the presence of corrupted measurements. Rather than discarding corrupted GPS data, our method retains and processes this information by running the Kalman filter with multiple observation models. 

Like the approach described in \cite{fan2017synchrophasor}, this paper aims to model sensor corruption as a parametric function, estimate the parameters, and then perform estimation with the corrected GPS measurements. However, our paper makes several significant contributions that further improve this idea. Unlike the discussed system, which deals with static systems, our approach performs online state estimation of a dynamical system. The SKF and parameter augmentation techniques identify the start of corruption, instead of relying on the previously discussed threshold-based methods which may be challenging to tune.

Our method requires hypothesizing a single (or several) potential parametric model for the GPS bias, where the corruption is characterized by specific parameters that define the nature and extent of the bias over time. 

The SKF is designed to operate across multiple potential observation models --- each one corresponding to an initiation of sensor fault --- and switch between them as it assesses the likelihood of each model being correct at any given time. This allows the system to pinpoint the exact time step at which the GPS data transitions from an uncorrupted state to a corrupted one. 

The model then identifies the parameters of the bias model using a parameter augmentation approach, where the bias parameters are treated as additional state variables within the Kalman filter. The filter simultaneously estimates the state of the system and the parameters of the corruption model, continuously refining these estimates as new data is processed. This dual estimation process ensures that both the system state and the nature of the GPS corruption are accurately tracked, enabling the recovery of the true system state even in the presence of significant GPS bias.

\subsection{Applications}
In this paper we demonstrate the effectiveness of the proposed methodology using two representative applications: balloon navigation in a moving atmosphere and shuttle reentry. These scenarios are representative of the significant challenges faced by existing state estimation methods when used with potentially corrupted GPS measurements.

\subsubsection{Balloon navigation} 
We investigate the use of our algorithm for online state estimation of a balloon drifting in a time-varying, 2-dimensional velocity field. The trajectory of a balloon was estimated using a set of external measurements corrupted by some bias. While this example does not involve
complex system dynamics, it demonstrates the performance of the proposed method and highlights its advantageous properties.

\subsubsection{Space shuttle reentry}
We explore the efficacy of using our algorithm on an application of  estimating position, velocity, and orientation of a space shuttle~\cite{betts2010practical}. Specifically, we estimate the state of the system IMU and a set of corrupted GPS measurements. This example demonstrates the performance of the method when used on GPS aided navigation systems.

\subsection{Organization of the paper}
The remainder of the paper is organized as follows: Section \ref{sec:background} provides the model background and problem setup, detailing the inertial navigation dynamics and the integration of GPS measurements. Section \ref{sec:gauss_filter} introduces the Gaussian filtering methodology and the specific implementation of the SKF. Section \ref{sec:switching_detection} describes the extention of SKF to corruption switching time detection. Section \ref{sec:results} discusses the application of the proposed method to two scenarios, balloon navigation and shuttle reentry, demonstrating its effectiveness. Finally, Section \ref{sec:conclusion} presents the conclusions and outlines potential future research directions.

\section{Model Background and Problem Setup}\label{sec:background}
This section provides background for inertial navigation.

\subsection{Frames}
We use two coordinate frames for describing the navigation dynamics: the body frame fixed to the vehicle and the Earth frame. The body frame is fixed to the aircraft center of mass and has orthogonal axes pointed forward along the longitudinal axis of the vehicle, downward, and rightward. We also assume that the IMU axes are aligned with the body frame axes, so that inertial readings are reported in body frame coordinates.

The Earth frame is fixed to the surface of the Earth and assumed to be inertial. It has axes pointing North, East, and downward. Throughout this paper we report the attitude, position, and velocity estimates with respect to this inertial frame. The superscripts $b$ and $i$ denote whether a given vector is resolved in the body or inertial frame, respectively. For example, the specific force $\vf^b\in\reals^3$ read by the accelerometers and the angular velocity $\ar\in\reals^3$ read by the gyroscopes are both resolved in the body frame.

\subsection{Inertial navigation}
In this section we describe inertial navigation dynamics. First, we present the navigation equations used to update position and orientation estimates using a direct measurements from accelerometers and gyroscope obtained from an IMU. Second, we present the error equations we use to correct the instrumentation drift introduced by the IMU. In, this paper we assume direct position measurements from a GPS unit for this task, however our approach can also be used with other external such as lidar or sonar in other settings \cite{tagliabue2021lion,michalec2014sidescan}.

\subsubsection{Navigation equations}\label{sec:IMU_update}
In this subsection, we follow the work of \cite{groves2015principles} to describe the process we use to update position and orientation estimates using the IMU readings. The process consists of a sequence of four steps to propagate the state variables through time: $\left. 1\right)$ update attitude; $\left. 2\right)$ transform specific force frame; $\left. 3\right)$ update velocity; $\left. 4\right)$ update position. To aid describing these four steps, we will use the $(-)$ notation to depict the initial states and $(+)$ to denote the updated states. However, before describing these steps in detail, we first explain the IMU model we use in this study. 

We consider a strapdown IMU sensors to provide biased and noisy measurements of specific force $\spf$ and angular velocity $\ar$, where specific force is the force sensed onboard while removing gravity. We denote the accelerometer and gyroscope biases of an IMU by  $\vb_a$ and $\vb_g$, respectively, and their corresponding noises by $\vn_a$ and $\vn_g$, respectively. Moreover, we assume that these sources of error enter the measurement additively, such that IMU readings $\spft$ and $\art$ satisfy
\begin{equation}\label{eq:biased_accel}
    \spft = \spf + \vb_a + \vn_a
\end{equation}
\begin{equation}\label{eq:biased_gyro}
    \art = \ar + \vb_g + \vn_g
\end{equation}
Understanding of these inertial navigation measurements is used to formulate a six degree of freedom model of the system’s dynamics. The state vector X of this model consists of the altitude, longitude, speed, flight path angle, azimuth, roll, pitch, yaw Euler angles, three component acceleration bias vector, and three component gyroscope bias vector --- resulting in a 15 state system:
\begin{equation}\label{eq:state}
    \vX =\begin{bmatrix}
        h & L & \lambda & v & \gamma & \alpha & \Phi & \Theta & \Psi & \vb_{a} & \vb_{g}
    \end{bmatrix},
\end{equation}
where Table~\ref{tab:notation} summarizes the notation used in this section.
\begin{table}[htbp]
    \centering
     \caption{Notation}
    \label{tab:notation}
    \begin{tabular}{|c|c|}
    \hline
    $g$                 &  Gravity\\
    \hline
    $h$                 &  Altitude \\
    \hline
    $L$                 &  Longitude\\
    \hline
    $\lambda$           &  Latitude\\
    \hline
    $v$                 &  Speed\\
    \hline
    $\vv$               &  Velocity\\
    \hline
    $\gamma$            &  Flight path angle\\
    \hline
    $\alpha$            &  Azimuth\\
    \hline
    $\ar$               &  Angular rate\\
    \hline
    $\spf$              &  Specific force\\
    \hline
    $\Phi$              &  Roll \\
    \hline
    $\Theta$            &  Pitch \\
    \hline
    $\Psi$              &  Yaw \\
    \hline
    \end{tabular} 
\end{table}
The inertial navigation equations update a state $\vX(-)$ into the next timestep using the IMU measurements. Rotational kinematics employ a forward-Euler type update to propagate the attitude angles with a timestep of $\Delta t$ according to
\begin{equation}\label{eq:euler_update}
    \begin{bmatrix}
        \Phi(+) \\ \Theta(+) \\ \Psi(+)
    \end{bmatrix} = \begin{bmatrix}
        \Phi(-) \\ \Theta(-) \\ \Psi(-)
    \end{bmatrix} + \begin{bmatrix}\dot{\Phi} \\ \dot{\Theta} \\ \dot{\Psi} \end{bmatrix} \Delta t.
\end{equation}
The Euler angular rates, denoted by a dot above the symbol, are obtained from a transformation of the angular velocity obtained from the gyroscope measurements according to
\begin{equation}\label{eq:euler_rates}
    \begin{bmatrix}\dot{\Phi} \\ \dot{\Theta} \\ \dot{\Psi} \end{bmatrix} = \begin{bmatrix}
        1 & \sin\Phi\tan\Theta & \cos\Phi\tan\Theta \\
        0 & \cos\Phi & -\sin\Phi\\
        0 & \sin\Phi/\cos\Theta & \cos\Phi/\cos\Theta
    \end{bmatrix}(\art - \vb_g).
\end{equation}
The translational dynamics follows a similar process. Specifically, given the coordinates of the velocity in the inertial frame, $\vv^i(-)$, the dynamics equations are given by
\begin{equation}\label{eq:velocity_update_nav}
    \vv^i \approx \vv^i(-) + \left[\vf^i + \vg^i(h(-))\right]\Delta t,
\end{equation}
where the specific force coordinates are transformed from the body frame  into the inertial frame
\begin{equation}\label{eq:spf}
    \vf^i \approx \frac{1}{2}(C^i_b(-) + C^i_b(+))(\spft - \vb_a)
\end{equation}
using the attitude matrix $C^i_b$ 
\begin{equation}\label{eq:321_euler}
    C^i_b = \begin{bmatrix}
    c\Theta c\Psi & c\Theta s\Psi & -s\Theta \\
    s\Phi s\Theta c\Psi - c\Phi s\Psi & s\Phi s\Theta s\Psi + c\Phi c\Psi & s\Phi c\Theta \\
    c\Phi s\Theta c\Psi + s\Phi s\Psi & c\Phi s\Theta s\Psi - s\Phi c\Psi & c\Phi c\Theta
    \end{bmatrix}.
\end{equation}
The gravity model we use provides the following gravity vector at altitude $\vg^i(h(-))$
\begin{equation}\label{eq:gravity}
    \begin{split}
        g_{E}^i(h) &= 0 \\
        g_{N}^i(h) &= 0 \\
        g_{D}^i(h) &= \frac{J_1}{(R_f + h)^2}
    \end{split}
\end{equation}
where the Gravitational constant $J_1 = 0.14076539\times10^{17}~\text{ft}^3\text{s}^{-2}$  the radius $R_f = 20,902,900$~ft~\cite{betts2010practical}, and the subscripts denote the East, North, and downward coordinates.

Next we convert the inertial-frame coordinates into the navigation-frame coordinates according to
\begin{equation}\label{eq:state_space_vel}
    \begin{split}
        v(+) &= |\vv^i| \\
        \gamma(+) &= asin\left(-\frac{v^i_D}{v}\right) \\
        \alpha(+) &= atan 2\left(\frac{v^i_E}{v^i_N}\right)
    \end{split}
\end{equation}
and update the positions
\begin{equation}\label{eq:position_update_nav}
    \begin{split}
      h(+)  &= h(-) - \frac{\Delta t}{2}\left(v_{D}^i(-) + v_{D}^i(+)\right) \\
      L(+) &= L(-) + \frac{\Delta t}{2}\left(\frac{v_{N}^i(-)}{R_f + h(-)} + \frac{v_{N}^i(+)}{R_f + h(+)}\right)\\
      \lambda(+) &= \lambda(-) + \frac{\Delta t}{2}\left(\frac{v_{E}^i(-)}{(R_f + h(-))\cos L(-)} \right. \\ &\left.+ \frac{v_{E}^i(+)}{(R_f + h(+))\cos L(+)}\right).
    \end{split}.
\end{equation}

Finally, the IMU biases are updated assuming random-walk dynamics, with Gaussian steps
\begin{equation}
    \begin{split}
        \vb_a(+) &= \vb_a(-) + \vn_a \quad\vn_a\sim\mathcal{N}(0,\Sigma_a) \\
        \vb_g(+) &= \vb_g(-) + \vn_g \quad\vn_g\sim\mathcal{N}(0,\Sigma_g) \\
    \end{split}.
\end{equation}

\subsubsection{Errors}
Mitigating the accumulation of IMU errors, requires an external information source. In this paper we use a linear observation model
\begin{equation}
Y_{k} = H(X_k) + \eta_{k-1},
\label{eq:meas_model}
\end{equation}
where $Y_{k} $ may represent, for example, GPS measurements at time step $k$, $H$ is the measurement function, and $\xi$ represents errors that arise from measurement noise. While more general considerations for $H$ are possible, we consider direct observations of the positions so that 
\begin{equation}
Y_{k} = \begin{bmatrix}
    h_{k} \\
    L_{k} \\
    \lambda_{k}
\end{bmatrix} + \eta_{k-1}.
\label{eq:GPS_meas}
\end{equation}

The primary objective of this study is to challenge the prevalent practice of discarding corrupted GPS measurementsby demonstrating the utility of identifying the temporal transition from an unbiased model, to a biased model when the GPS data becomes corrupted. With this goal, we accounted for possible measurement corruption, by replacing the measurement model $H$ from Eq.~\eqref{eq:meas_model} with the function
\begin{equation}\label{eq:obs_switch}
  H(X_k,\theta, t_s, t_k) = 
    \begin{cases}
      \nomobs(X_k), & \text{if }t_k \leq t_s, \\
      \corobs(X_k,\theta, t_s, t), & \text{if } t_k > t_s,
    \end{cases},
\end{equation}
where $t_s$ denotes the time at which the observation model, parameterized by $\theta$, switches modalities or becomes corrupted. Additionally, the variables $s$ and $\theta$ are uncertain parameters specifying the corrupted observation model. 

More specifically, $s$ indexes the timestep $t_s$ at which there is a switch from the nominal observation model $\nomobs:\reals^{\dimx}\mapsto\reals^{\dimy}$ to its corrupted counterpart $\corobs:\reals^{\dimx}\times\reals^{\dimth}\times\reals\times\reals\mapsto\reals^{\dimy}$, and $\theta$ is a collection of parameters that define $\corobs$. 

To represent the uncertainty in these variables, we model $\theta$ as a $\dimth$-dimensional continuous random vector and $s$ a discrete random variable, which takes values in $\{1,\ldots,n\}$, with $n$ being the maximum number of timesteps. Finally, to allow our model to function without any knowledge on when the switch will occur, we specify a uniform prior distribution over $s$ such that its probability mass function is defined as $p(s)=1/n$.

In essence, the corrupted observation model, $\corobs$  represents the combination of unbiased, nominal model $\bar{H}$ with the added bias function, $H_b$. Some example bias functions that are considered later in the paper, are provided in Table \ref{tab:bias_functions}.

\begin{table}[ht!]
\caption{Bias models with corresponding functions and parameters}
\begin{center}
\begin{tabular}{|l|l|l|}
\hline
    Model & Equation & Parameters \\
    \hline
    Static & $H_b(\theta,t_s,t_k) = A$ & $\theta = [A]$ \\
    \hline
    Linear & $H_b(\theta,t_s,t_k) = A + B(t_k-t_s)$ & $\theta = [A,B]$ \\
    \hline
    Quadratic & $\!\begin{aligned}[t]
        H_b(\theta,t_s,t_k) &=A + B(t_k-t_s) \\
        &+ C(t_k-t_s)^2 
    \end{aligned}$ & $\theta = [A,B,C]$\\
\hline
\end{tabular}
\label{tab:bias_functions}
\end{center}
\end{table}

\section{Gaussian filtering}\label{sec:gauss_filter}
In this section we describe the Gaussian filtering methodology that we use to fuse the observations $Y_k$ with the dynamics to obtain more accurate state estimates. The joint dynamics and observation equations are coupled according to
\begin{subequations}
  \begin{align}
    X_{k} &= \Phi(X_{k-1}) + \xi_{k-1}, &\quad \xi_{k-1}\sim\mathcal{N}(0,\Sigma) \label{eq:dyn_nav}\\
    Y_{k} &= H(X_k) + \eta_{k-1}, &\quad \eta_{k-1}\sim\mathcal{N}(0,\Gamma) \label{eq:obs_nav}
  \end{align}
\end{subequations}
where $k\in [\numy]$ is a time index. The states and measurements at time $t_k$ are denoted by $X_k$ and $ Y_k$; $\Phi$ and $H$ are the dynamics and measurement models; and finally $\xi_{k}$ and $\eta_{k}$ correspond to the Gaussian process and measurement noises with zero mean and respective covariances $\Sigma$ and $\Gamma$. 

Next, as standard practice, we approximate the filtering distribution \(p(X_k|\mathcal{Y}_k)\) as a Gaussian. $p(x_k| \mathcal{Y}_k) = \mathcal{N}(m_k, C_k)$, where $\mathcal{Y}_k = \{ Y_1,\ldots, Y_k \}$ is the observation history, $m_k$ is the mean of the state estimate at time step $k$, and $C_k$ is the covariance matrix. Furthermore, we  compute marginal log likelihood, $\mathcal{L}$ As
\begin{equation}
    \mathcal{L} = \frac{1}{2}\left(-log(|D_k|) - (\mathcal{Y}_k-\mu_k)^{T}D_k^{-1}(\mathcal{Y}_k-\mu_k)-n\log{2\pi}\right)
\end{equation}
where $\mu_k$ is the predicted mean of the observation, $D_k$ is the covariance of the predicted observation, and n is the dimensionality of the observations. We use the marginal likelihood to access how well the data aligns with model predictions. In particular, we use it to compare how well different filters match the data, to select the most likely switching branch. Since the log marginal likelihood is used for the comparison, we can simplify this expression by removing the common constants to
\begin{equation}
    \mathcal{L} = -log(|D_k|) - (\mathcal{Y}_k-\mu_k)^{T}D_k^{-1}(\mathcal{Y}_k-\mu_k)
\end{equation}
In this paper, we use the specific version of Gaussian filter, namely the Unscented Kalman Filter (UKF). However, the methodology described below does not depend specifically on the UKF and is applicable to any Gaussian Filter, and so we do not provide these details here. The explicit prediction and update steps can be found in~\cite{sarkka2023bayesian}. 

\subsection{Filtering challenges for uncertain observation models}
Gaussian filtering can be used  to compute the distribution $p(X_k | \yk)$ when all parameters of the observation model in Eq. (\ref{eq:obs_switch}) are known. However, when $\theta$ and $s$ are uncertain, evaluating the filtering distribution $p(X_k |\yk)$ requires marginalizing over $\theta$ and $s$ through the computation
\begin{equation}\label{eq:marginalize}
  p(X_k | \yk) = \sum_{s\in\K}\int_{\reals^{\dimth}}p(X_k,\theta,s|\yk)\rmd \theta.
\end{equation}
This marginalization integrates over all the parameters and time indices $s$ simultaneously --- a computationally prohibitive procedure for many problems. However, by factoring the joint distribution, Eq.~\ref{eq:marginalize} can be expressed as a probability-weighted average of the marginals over the parameters
\begin{equation}\label{eq:factorized}
  p(X_k | \yk) = \sum_{s\in\K}p(s|\yk)\int_{\reals^{\dimth}}p(X_k,\theta|s,\yk)\rmd \theta.
\end{equation}
Here, the weights $p(s | \yk)$ represent the probability of a switch at time $s$ given the data, and the entire summation can be viewed as a model averaging procedure. Moreover, both the integral and the summation of Eq.(\ref{eq:factorized}) can be efficiently approximated through the methodology discussed in the next section.

\subsection{Switching Kalman Filter}
The computational cost of keeping and assessing the probability of a switching time at each time-step grows exponentially with time. Consequently, we develop a variation of switching Kalman filter \cite{murphy1998switching} to reduce this growth in computational expense. The SKF is typically used to select the most probable mode between several possible modes of operation, and we employ it here for the task of identifying the observation model switching time.

Classically, this problem is closely related to the task of considering multiple modes of operation. Let $\Phi$ and $H$ take in a second variable $S$ denoting the mode of operation:
\begin{align}
    X_{k} &= \Phi(X_{k-1}, S_{k-1} ) + \xi \quad \xi \sim \mathcal{N}(0, \Sigma) \\
    Y_{k} &= H(X_k, S_k) + \eta \quad \eta \sim \mathcal{N}(0, \Gamma)
\end{align}
where we have abused notation to denote that $H(X_k, S_k)$ is a linear model in $X_k$ that may depend non-linearly on $S_k$. 

The switching Kalman filter involves the simultaneous operation of multiple filters, each predicated on a distinct assumption of $M_k$ within the observational and dynamics models. The comparative evaluation of these filters is facilitated through the calculation of the marginal likelihoods enabling the identification of the most probable mode of operation at each time step. These likelihoods are easily computed due to the Gaussianity assumption of the state estimate for each mode of operation.

\begin{figure}[ht!]
    \centering
    \includegraphics[width=0.5\textwidth]{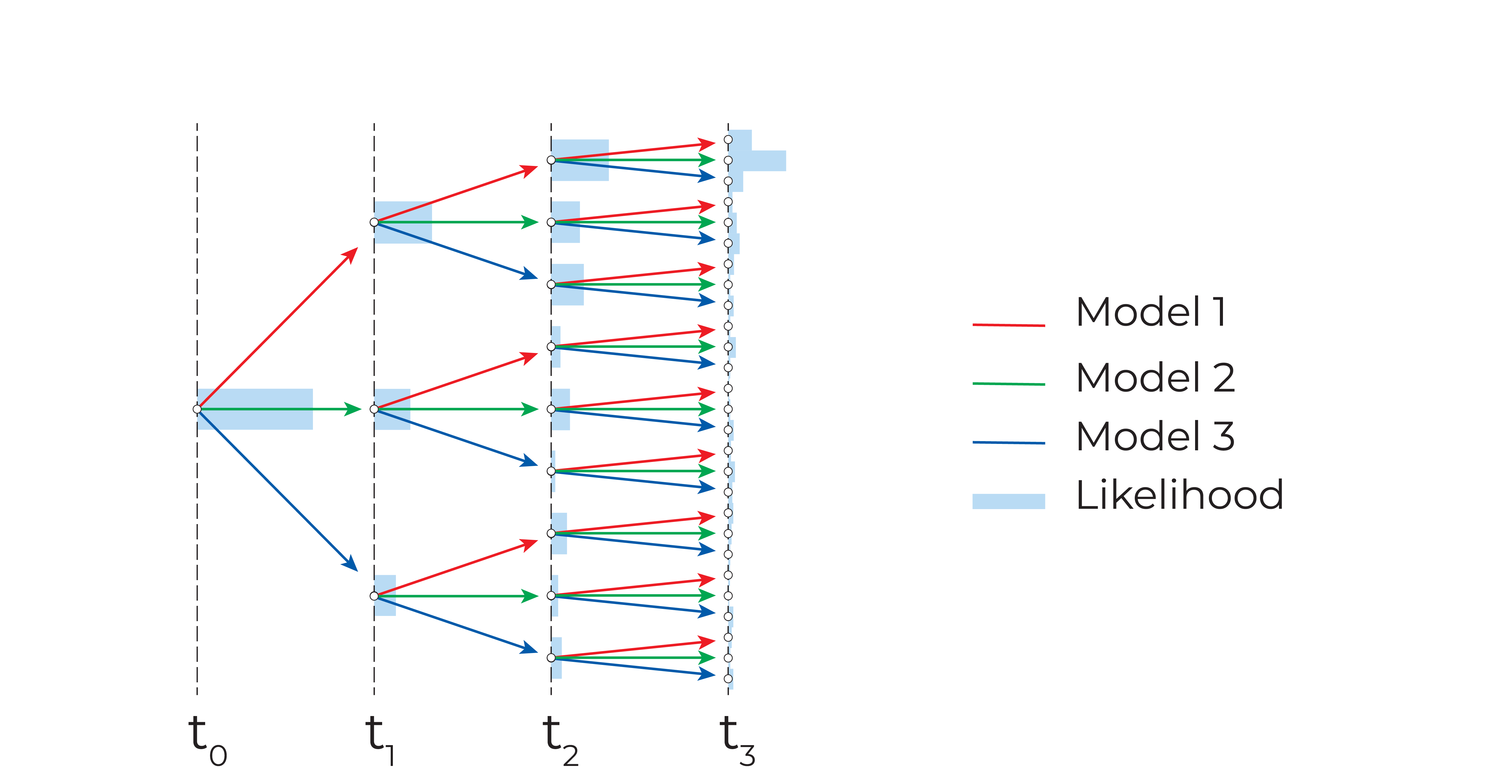}
    \caption{Diagram representing the traditional SKF operational principle. Red, green, and blue arrows represent the use of model 1, 2, or 3 respectively as an observation model. At every single instance can switch to using any of the three models. The length (left-to-right) of the blue likelihood bars represent the size of the likelihood.}
    \label{fig:traditional_SKF}
\end{figure}

The operational principle of employing SKF is illustrated in Figure~\ref{fig:traditional_SKF} for a conceptual example, not tied to any specific dynamical system. The figure demonstrates that, after three timesteps, the likelihood for the filter utilizing the first mode of operation for the first two time steps and the second mode for the third time step, is greater than that of other scenarios. This suggests that after initially assimilating data using the first mode of operation we should switch to using the second mode of operation for the third, and all subsequent time steps.

As shown in Figure~\ref{fig:traditional_SKF}, the number of potential combinations of operating modes grows exponentially with time. For example, when using $M$ distinct models, $M^t$ Kalman filters must run in parallel at time $t$. However, collapsing, selection, iterative, and variational approaches
\cite{murphy1998switching} can be used to reduce the computational cost of SKF. In Section \ref{sec:switching_detection} we describe a variation of selection approach we developed to effectively employ SKF for detecting the time at which sensor corruption is initiated. 

\section{Application to switching time detection}\label{sec:switching_detection}

In this section, we present an approach to estimate the state of the system under sensor manipulation. First, we introduce the parameter augmentation approach to learn the parameters of the corruption model. Second, we describe the modifications made to the traditional SKF that we use to apply it for identifying the corruption switching time. Third, we present further refinement of the SKF formulation to reduce computational cost. 

For each model choice we employ a parameter augmentation methodology. Specifically, we integrate the parameters of the bias model into the state vector, subsequently estimating them alongside dynamic states through prediction and update procedures of the Gaussian filter. The parameter-state augmentation approach essentially filters the following equations for the augmented state  $\bar{X}_k = [X_k, \theta]$
 \begin{align}
      \bar{X}_{k} &= \bar{\Phi}(\bar{X}_{k-1} ) + \xi_k &\quad \xi_k \sim \mathcal{N}(0, \bar{\Sigma}) \label{eq:dyn_nav_aug}\\
      Y_{k} &= \bar{H}(\bar{X}_k, t_s, t_k) + \eta_{k}, &\quad \eta_{k}\sim\mathcal{N}(0,\Gamma) \label{eq:obs_nav_aug}
\end{align}
where $\bar{\Phi}$ and $\bar{H}$ augment the original dynamics with the identity corresponding to unchanging parameters. Similarly, $\bar{\Sigma}$ now includes the process noise covariance for the random walk dynamics of the parameters.

\begin{figure}[ht!]
    \centering
    \includegraphics[width=0.5\textwidth]{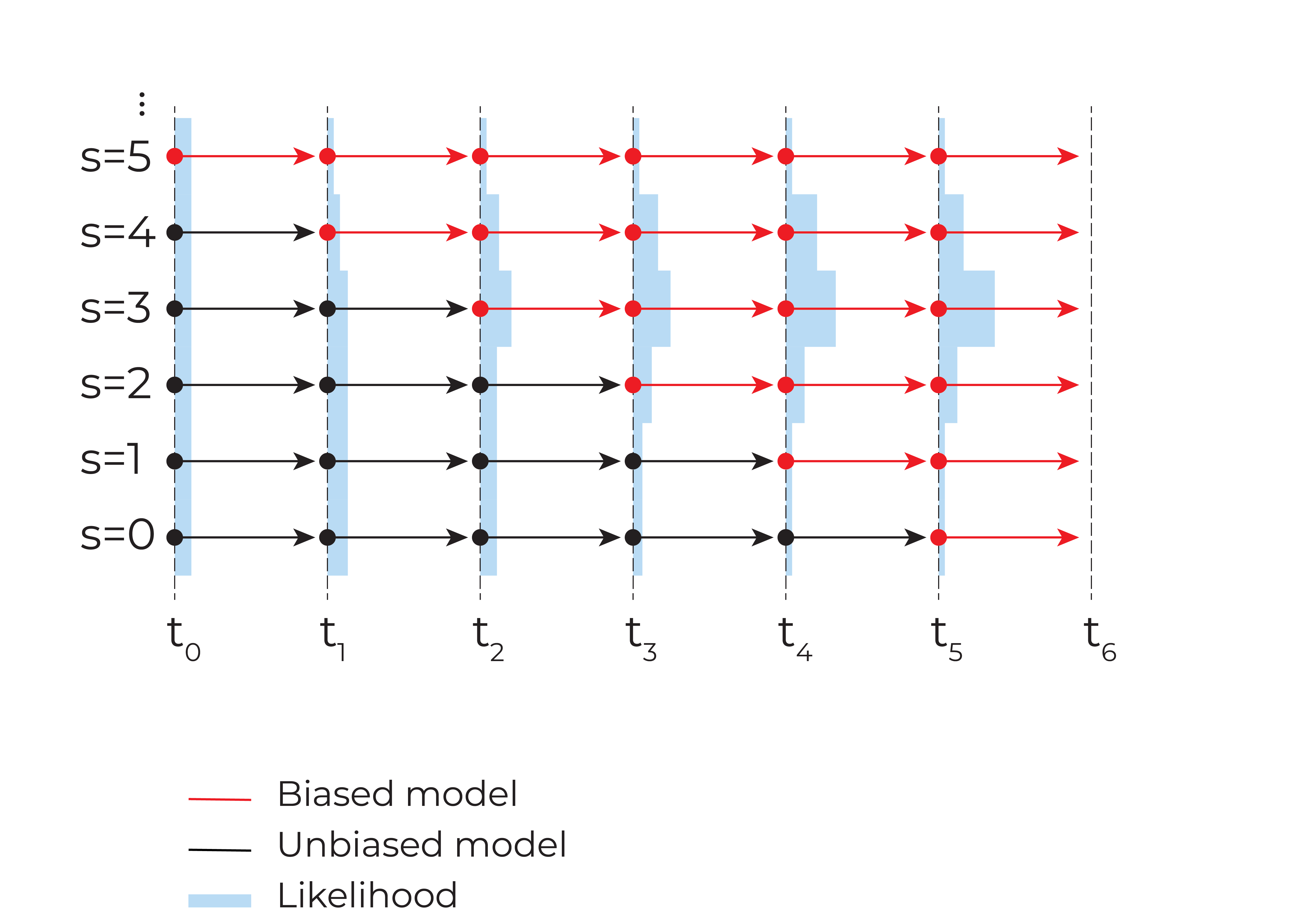}
    \caption{Diagram representing the parallel SKF operational principle. Red dots indicate the usage of a biased observation model, whereas black dots indicate an unbiased model. As an example $S=3$ indicates that the observation model is corrupted at $t=2$. The length (left-to-right) of the blue likelihood bars indicates the size of the likelihood.}
    \label{fig:parallel_SKF}
\end{figure}

Next, we adopt the switching Kalman filter to the task of identifying the switching time by setting $S_k$ to the switching time, $t_s$ and then simultaneously run a filter for each possible switching time. This allows us to accurately estimate the switching time, with a precision up to half a time step. The conceptual operation of the resulting parallel SKF is illustrated in Figure~\ref{fig:parallel_SKF}, using an artificial example not tied to any specific dynamical system. The figure demonstrates that, after six time steps, the likelihood for the filter utilizing $t_2$ as a switching time surpasses that of other scenarios, suggesting an approximation that the bias began influencing the observations at $t_2$.

Unfortunately, running a filter for each possible switching time is computationally demanding. Consequently, we reduce the computational cost of evaluating the filtering distribution $p(X_k|\yk)$, by approximating the summation from Eq.~\eqref{eq:factorized} in an efficient manner. The form of our approximation was based on the following two observations: 
\begin{enumerate}
\item The nested nature of the different models, i.e. models $s=i$ and $s=j$ are equivalent up until time $t_{\min\{i,j\}}$.
\item The number of branches is reduced by retainining only those with the highest likelihood 
\end{enumerate}

\begin{figure}[ht!]
    \centering
    \includegraphics[width=0.5\textwidth]{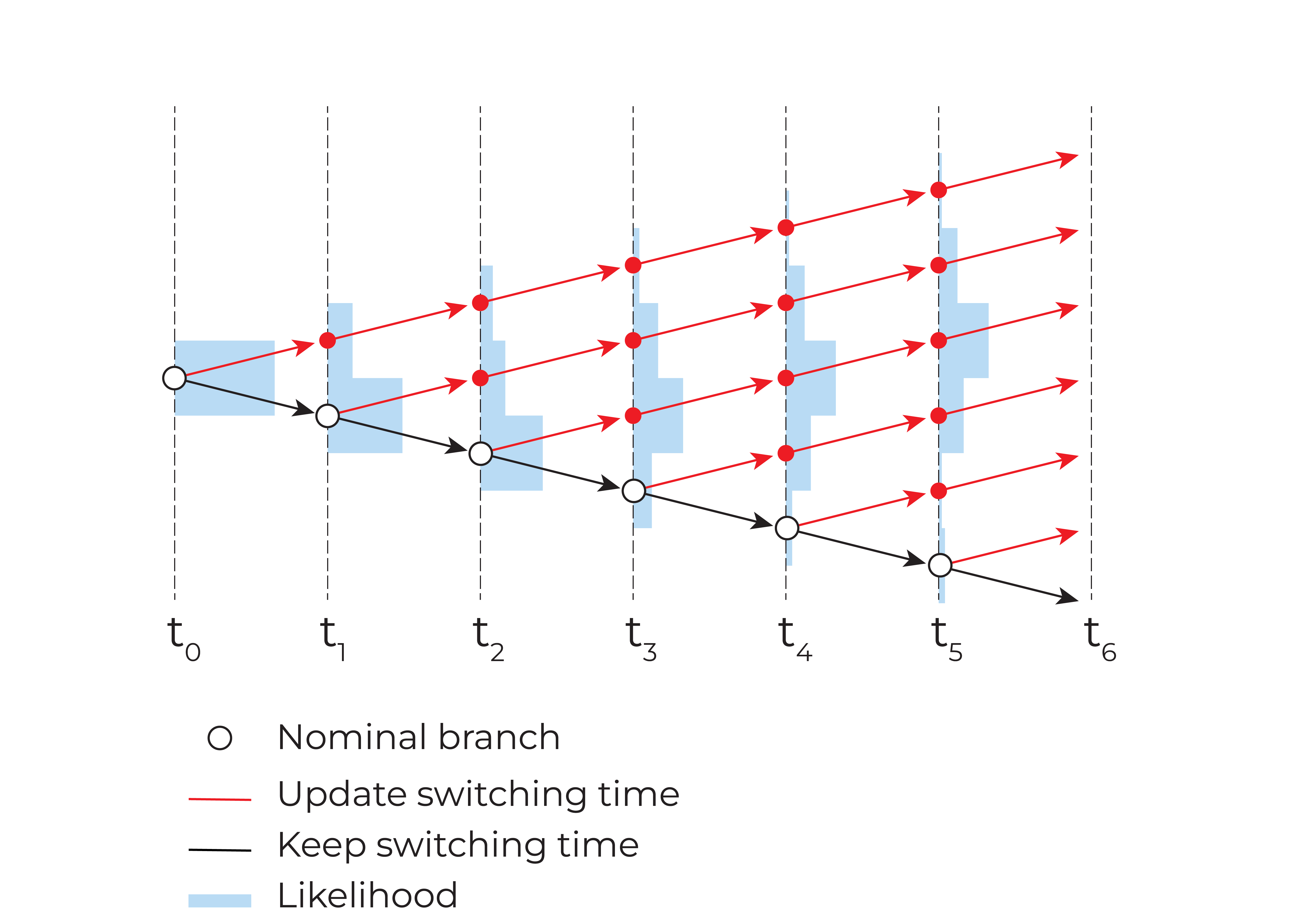}
    \caption{Diagram representing the branched SKF operational principle. Red dots indicate the usage of a biased observation model, whereas hollow dots indicate an unbiased model. At every instance a new biased branch starts from the unbiased one. The length (left-to-right) of the blue likelihood bars indicates the size of the likelihood.}
    \label{fig:branched_SKF}
\end{figure}

 We use the first observation to avoid the redundant operation of multiple filters yielding identical outcomes. Specifically, we only run a singular filter model, predicated on an entirely unbiased observational model (hereafter referred to as the nominal branch) up to each switching time. Then at each switching time we initialize a new filter (which we call corrupted branches) that uses a biased model from that point in time. This process, henceforth known as branched SKF, is conceptually depicted in Figure~\ref{fig:branched_SKF}, and substantially reduces the computational cost of running parallel SKF, while maintaining its performance. 

To further reduce computational cost, we prune the corrupted branches by eliminating those with the lowest likelihood. This procedure involves computing the marginal likelihood of each corrupted branch at each time step and retaining only the $M$ filters associated with the $M$ highest likelihoods. Note that the nominal branch is exempt from discarding irrespective of its likelihood and the size of $M$ must be determined by the computational resources available. 

\begin{figure}[ht!]
    \centering
    \includegraphics[width=0.5\textwidth]{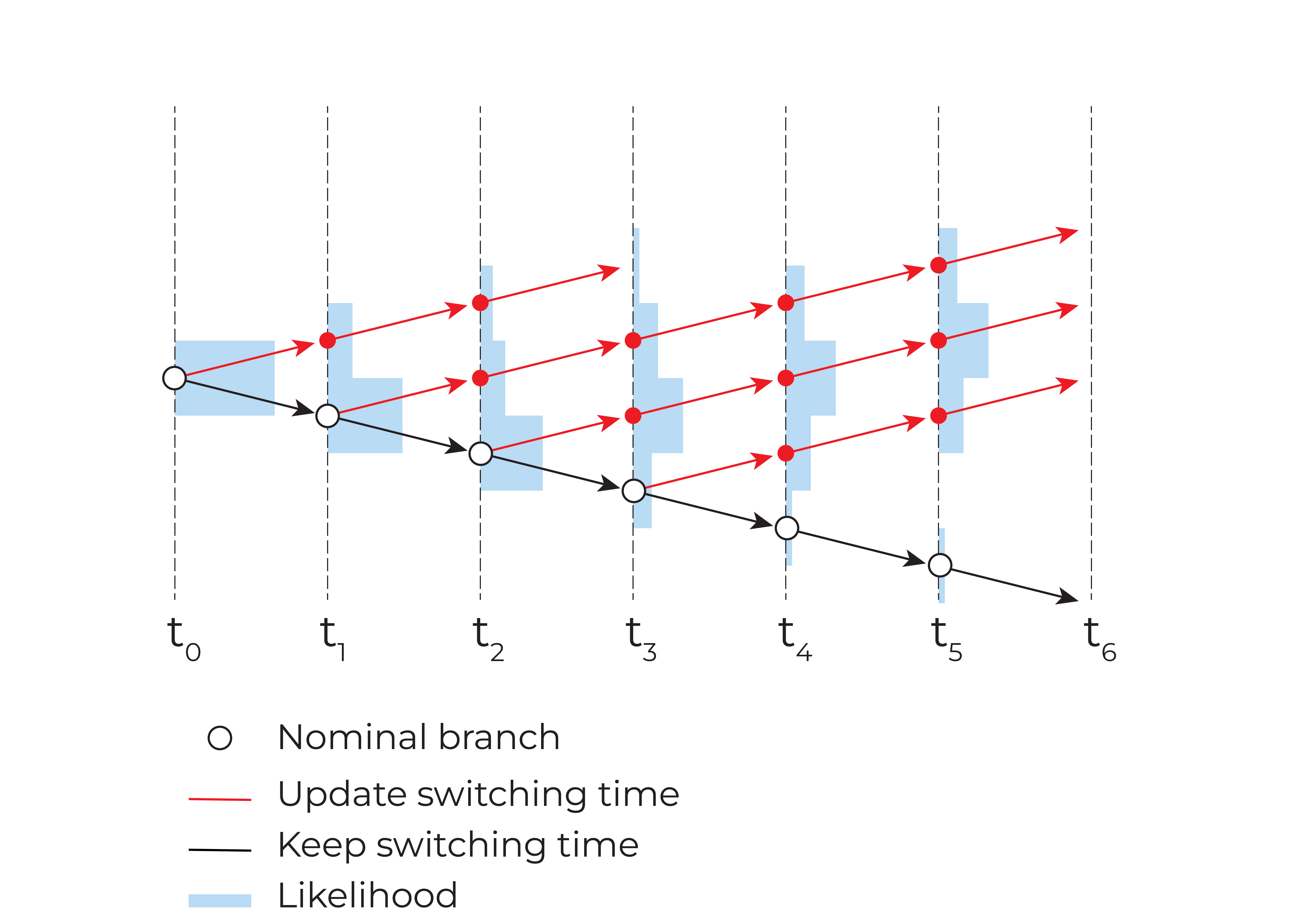}
    \caption{Diagram representing the selective branched SKF operational principle. Red dots indicate the usage of a biased observation model, whereas hollow dots indicate an unbiased model. At every instance a new biased branch starts from the unbiased one. Once there are more than three corrupted branches, the less likely one is removed. As an example, at $t_3$ the branch corrupted at $t_0$ is removed. The length (left-to-right) of the blue likelihood bars indicates the size of the likelihood.}
    \label{fig:eff_SKF}
\end{figure}

Figure~\ref{fig:eff_SKF} depicts the conceptual process that selective branched SKF approach employs for the case of maintaining only the three most probable branches ($M = 3$). We observe that by step $t_3$, the branch predicated on an initially unbiased observational model is deemed to have a significantly lower likelihood, leading to its exclusion from further computations. Similarly, branches assuming the switching time of $t_4$ and subsequent steps are discarded based on their comparatively lower likelihood against filters operating under a switching time closely matching the actual switching time of bias, as exemplified by $t_2$ in this instance.

Algorithm \ref{alg:skf_alg} summarizes the main steps of the branched SKF process. Step 1 initializes the unbiased branch $F_0$. The unbiased branch, along with all biased branches that are created later in the algorithm are characterized by the time used for switching $(t)$, the likelihood of the branch $(\mathcal{L})$, an array of estimated states $(m)$, and an array of covariance matrices $(C)$. When initializing the unbiased branch, step 1 sets $t$ and $\mathcal{L}$ to zero, the first entry of $m$ to the initial condition, and the remaining entries to zero vectors with matching dimension. The first entry of $C$ is set to the initial covariance, while the remaining entries are set to identity matrices of the appropriate size. Step 2 then initializes the set of all branches, denoted as $F$, which initially contains only the unbiased branch. Step 3 loops over all time steps $k=1, \dots ,N$. At each timestep, step 4 updates the current time $t$ and step 5 makes predictions for all branches in $F$ using Algorithm \ref{alg:pred_alg}. If data is available Step 7 then updates the unbiased branch using the unbiased observation model $\bar{H}$. Steps 8-9 then update all corrupted branches using the corrupted observation model $\tilde{H}$. Finally steps 11-13 identify the M-1 corrupted branches that have the highest likelihood and removes the other branches from the set F using Algorithm \ref{alg:clean_alg}. It is important to note, that the nominal branch updates twice. First, we update the nominal branch using the unbiased observation model in step 7 to maintain one branch free from corruption. Second, in steps 8-9 the nominal branch is updated again using corrupted observation model which represents the corruption initiated at the current timestep. 

\begin{algorithm}
    \caption{Efficient SKF to learn $t_0$}\label{alg:skf_alg}
    \textbf{Inputs:}\\
    \hspace*{\algorithmicindent}\text{Number of time steps $N$}\\
    \hspace*{\algorithmicindent}\text{Sampling frequency $\delta$}\\
    \hspace*{\algorithmicindent}\text{Process noise covariance $Q$}\\
    \hspace*{\algorithmicindent}\text{Measurement noise covariance $R$}\\
    \hspace*{\algorithmicindent}\text{Initial state $x_0$}\\
    \hspace*{\algorithmicindent}\text{Initial covariance $C_0$}\\
    \hspace*{\algorithmicindent}\text{Time step $\Delta t$}\\
    \hspace*{\algorithmicindent}\text{GPS measurements $\mathcal{Y}$}\\
    \hspace*{\algorithmicindent}\text{Number of filters $M$}\\
    \textbf{Outputs:} \\
    \hspace*{\algorithmicindent}\text{A set of most likely branches $F$}
    \begin{algorithmic}[1]
        \State $F_0 = [0, 0, [x_0,0,\dots,0 ],  [C_0,I,\dots,I ] ]$ initialize the unbiased branch as $[t, \mathcal{L}, m, C ]$
        \State $F = \{F_0\}$ set of all branches
        \For{\texttt{$k$ in $1,2,\dots,N$}}
            \State $t = k\Delta t$; update current time
            \State $F = $Predict($F, Q$); prediction for all branches
            \If{$k \% \delta$ = 0}
                \State $F_{new} = \{$Update($F_0, \bar{H},\mathcal{Y}_k, R)\}$; update unbiased branch with unbiased observation model
                \For{$F_i \in F$}
                    \State $F_{new}$.append(Update($F_i, \tilde{H},\mathcal{Y}_k, R$)); update all branches with corrupted observation model
                \EndFor
                \State F = $F_{new}$; update set of branches
                \If{len($F$) $> M$}
                    \State $F =$ Clean($F$)
                \EndIf
            \EndIf             
        \EndFor
    \end{algorithmic}
\end{algorithm}

Algorithm \ref{alg:pred_alg} summarizes the prediction step of an Algorithm \ref{alg:skf_alg}. It takes the set of all branches and process noise covariance matrix $Q$, and then propagates each branch $F_i$ through the standard prediction step of the chosen Gaussian filter.

\begin{algorithm}
    \caption{Predict}\label{alg:pred_alg}
    \textbf{Inputs:}\\
    \hspace*{\algorithmicindent}\text{Set of all branches $F$}\\
    \hspace*{\algorithmicindent}\text{Process noise covariance $Q$}\\
    \textbf{Outputs:} \\
    \hspace*{\algorithmicindent}\text{Set of all branches $F$}
    \begin{algorithmic}[1]
        \For{$F_i \in F$}
            \State $t_s, \mathcal{L}, m, C = F_i$
            \State $m_{k+1}, C_{k+1} = \texttt{predict}(m_k,C_k,Q)$;
            \State $F_i = [t_s, \mathcal{L}, m, C]$
        \EndFor
    \end{algorithmic}
\end{algorithm}

Algorithm \ref{Update_alg} summarizes the update step described by steps 7-9 of an Algorithm \ref{alg:skf_alg}. It takes a branch $F_i$, an observation model $H$, new observation $\mathcal{Y}_k$, and a process noise covariance matrix $R$ from the current step $k$, and propagates the branch through the standard update step of the chosen Gaussian filter using the provided observation model. The algorithm then updates the marginal likelihood $\mathcal{L}$ of the branch and returns the updated branch to the main algorithm.

\begin{algorithm}
    \caption{Update}\label{Update_alg}
    \textbf{Inputs:}\\
    \hspace*{\algorithmicindent}\text{Branch $F_i$}\\
    \hspace*{\algorithmicindent}\text{Observation model $H$}\\
    \hspace*{\algorithmicindent}\text{New observations $\mathcal{Y}_k$}\\
    \hspace*{\algorithmicindent}\text{Process noise covariance $R$}\\
    \textbf{Outputs:} \\
    \hspace*{\algorithmicindent}\text{Branch $F_i$}
    \begin{algorithmic}[1]
        \State $t_s, \mathcal{L}, m, C = F_i$
        \State $\displaystyle{m_{k+1}, C_{k+1}, \mu, D = \texttt{Update}(m_{k+1},C_{k+1}, \mathcal{Y}_k, t_s, H)}$
        \State $\mathcal{L} = \mathcal{L}-\log(|D|) - (\mathcal{Y}_k-\mu)D^{-1}(\mathcal{Y}_k-\mu)$
        \State $F_i = [t_s, \mathcal{L}, m, C]$
    \end{algorithmic}
\end{algorithm}

Algorithm \ref{alg:clean_alg} removes the branch with the lowest likelihood. It takes in the set of all branches, identifies the greatest negative marginal likelihood,  removes the corresponding branch, and then returns the reduced set of branches back to the main algorithm.

\begin{algorithm}
    \caption{Clean}\label{alg:clean_alg}
    \textbf{Inputs:}\\
    \hspace*{\algorithmicindent}\text{Set of all branches $F$}\\
    \textbf{Outputs:} \\
    \hspace*{\algorithmicindent}\text{Reduced set of all branches $F_{new}$}
    \begin{algorithmic}[1]
        \State $L = \{\mathcal{L}_1, \dots \mathcal{L}_{len(F)}\}$
        \State $i_{max} = \texttt{argmax}(L)$
        \State $F_{new}  = F - F_{i_{max}}$
    \end{algorithmic}
\end{algorithm}

\section{Numerical Experiments}\label{sec:results}
In this section, we present the results of a numerical study investigating the properties of our novel state estimator on two examples. The first example models a balloon traversing a velocity field to demonstrate that SKF can accurately capture the switching time under different scenarios. The second example, demonstrates the performance of SKF when using an IMU as the basis of a simple shuttle reentry model. For both examples, the true corruption models will encompass the Static, Linear, and Quadratic bias models as defined in Table \ref{tab:bias_functions} to test the sensitivity of our SKF to different corruption models. However, when learning, we always assume the corruption to be quadratic and learn the parameters of this quadratic. In other words, at each timestep we learn assuming the most complex corruption model. However, if the observational models were fundamentally different, our SKF could be used to consider multiple observational models at the same time. 

In this study we use the error in the corruption time identified by the SKF as the primary measure of performance. However, we also measure the accuracy of the estimated trajectory relative to the true one using the relative Root Mean Squared Error -- $RMSE$\
\begin{equation}
    RMSE = \sqrt{\frac{\sum_{k=1}^{N}\left(m_k - X_k\right)^2}{\sum_{k=1}^{N}X_k^2}}
    \label{eq:rmse}
\end{equation}
where  $m_k$ denotes the estimated state, and $X_k$ denotes corresponding true trajectory  at time step $k$.
\subsection{Drifting Balloon}

The first example we address in this paper is the joint parameter-state estimation problem of a balloon navigating within a discretized, time varying, 2-dimensional velocity field over longitude and latitude.

\subsubsection{Problem Formulation}

\begin{figure}[ht!]
    \centering
    \includegraphics[width=0.49\textwidth]{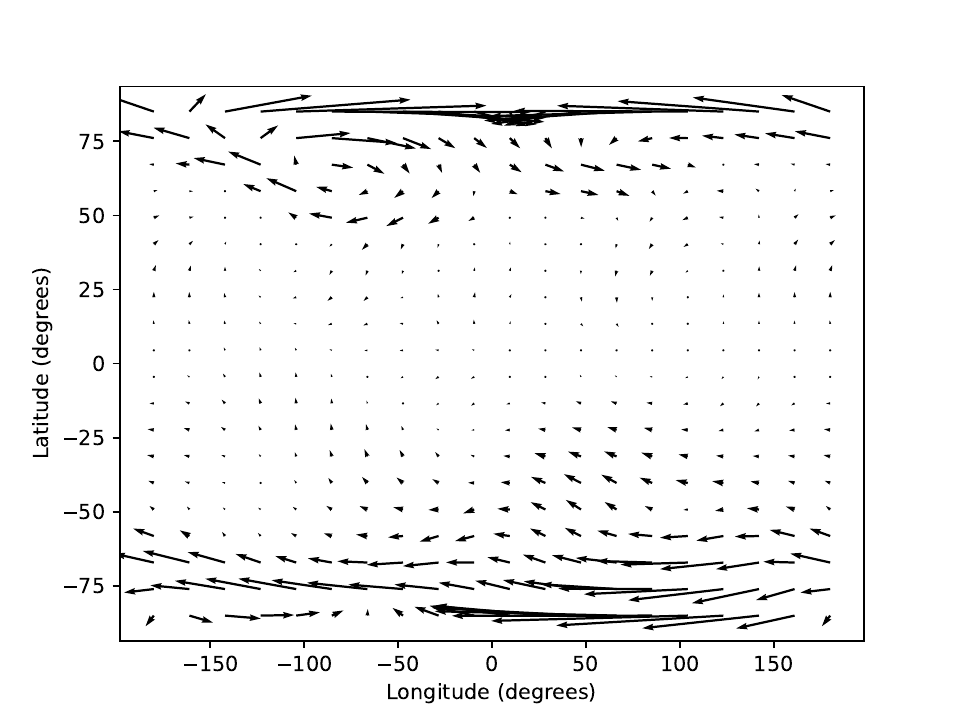}
    \caption{The velocity field used for the balloon drifting example problem at the initial time step}
    \label{fig:velocity_field}
\end{figure}

For the purposes of this analysis, the balloon is assumed to be a point mass and its initial position is assumed to be known. The dynamics are then simplified according to Eq. (\ref{eq:lonlat},\ref{eq:balloon_dyn},\ref{eq:balloon_obs}). Here, the state $X$ of the balloon is described by its longitude and latitude measured in degrees that are expressed as $x_{lon}$ and $x_{lat}$, respectively; $\Delta t$ is the length of a single time step set to $0.01$ hours;  $h_b$ is a true bias function that was used to corrupt the measurements in the data generation process;  and finally, $Q_{k}$ and $R_{k}$ represent process and measurement noise covariance matrices, respectively. Values $u(X, t)$ and $v(X, t)$ respectively describe the longitude and latitude velocities of the selected velocity field at the location of the balloon at the time $t$. The HWM14 model~\cite{Drob:2008,Drob:2015} is used to generate a notional, time varyiung velocity field, and the initial state is depicted in Figure~\ref{fig:velocity_field}. 

\begin{equation}\label{eq:lonlat}
X = \begin{bmatrix}
    x_{lon} \\
    x_{lat}
\end{bmatrix}
\end{equation}
\begin{equation}\label{eq:balloon_dyn}
    X_{k+1} = X_{k} + \Delta t \begin{bmatrix}
        u(X_{k}, t_k) \\
        v(X_{k}, t_k)
    \end{bmatrix} + \xi_{k}\quad\xi_{k}\sim\mathcal{N}(0,Q_{k})
\end{equation}
\begin{equation}\label{eq:balloon_obs}
    Y_{k} = X_{k}+H_b(\theta,t, t_s)+ \eta_{k}\quad\eta_{k}\sim\mathcal{N}(0,R_{k})
\end{equation}

The trajectory of a balloon system is simulated for 5 hours (500 time steps) with the initial position of -35 degrees longitude and 25 degrees latitude. The process and measurement noise covariance matrices remained constant throughout the simulation and are defined according to 
\begin{align}
    Q_k &= \begin{bmatrix}
        q_x & 0 \\
        0 & q_x \\
    \end{bmatrix}
    \label{eq:qx} \\
    R_k &= \begin{bmatrix}
        r & 0 \\
        0 & r \\
    \end{bmatrix}
    \label{eq:rx}
\end{align}
where $q_x$ is the process noise variance and $r$ is the measurement noise variance. During parameter augmentation, the process noise matrix $Q_k$ is augmented with the parameter process noise with variance $q_p$ that is kept identical for static, linear, and quadratic bias. The process noise for the parameter-state augmented system is 
\begin{equation}
    Q_{k,aug} = \begin{bmatrix}
        q_x & 0 & 0 & 0 & 0 \\
        0 & q_x & 0 & 0 & 0 \\
        0 & 0 & q_p & 0 & 0 \\
        0 & 0 & 0 & q_p & 0 \\
        0 & 0 & 0 & 0 & q_p \\
    \end{bmatrix}
    \label{eq:qx_aug}
\end{equation}
The initial covariance matrix for the filter is set to $I_5$, the initial filter mean is set to the initial condition for the states and to zero for the augmented parameters.

In the next two subsections we analyze how the bias parameters, switching times, and filtering setup affect the ability to identify the corruption onset and perform state estimation. We first provide and describe a set of representative behaviors in Section~\ref{sec:representative_balloon} and then provide a statistical analysis in Section~\ref{sec:statistical_balloon}.

\subsubsection{Representative examples}\label{sec:representative_balloon}
In this section we provide
some representative behaviors of the algorithm's performance in response to different problem setups. Table \ref{tab:tests_balloon} summarizes the test setup parameters and results for these experiments. Each test is color-coded based on the performance. Cases where the switching time is identified up to a single time step are highlighted in green; yellow denotes the experiments where the switching time is identified correctly within 10 time steps; and red highlights the experiments where the SKF is not able to retrieve switching time at all. Note, that in the unbiased cases (e.g., Test 8),
the branches that survive the entire simulation, all start at the very end and have very similar marginal likelihoods, including the unbiased branch. Therefore, any result close to the end of the simulation is considered to be a success and highlighted in green.

\begin{table*}[htbp]
    \centering
    \caption{Representative test cases for the drifting balloon}
    \label{tab:tests_balloon}
    \resizebox{\textwidth}{!}{
    \begin{tabular}{|c|cccccccccc|}
        \arrayrulecolor{white}\hline
        \rowcolor{Gray}
        & \multicolumn{1}{c|}{Measurement} & \multicolumn{1}{c|}{Process} & \multicolumn{1}{c|}{Sampling} & \multicolumn{1}{c|}{Static} & \multicolumn{1}{c|}{Linear} & \multicolumn{1}{c|}{Quadratic} & \multicolumn{1}{c|}{True} & \multicolumn{1}{c|}{Estimated} & \multicolumn{1}{c|}{Latitude} & \multicolumn{1}{c|}{Longitude}\\
        \rowcolor{Gray}
        \multirow{-2}{*}& \multicolumn{1}{c|}{Noise Variance} & \multicolumn{1}{c|}{Noise Variance} & \multicolumn{1}{c|}{Frequency} & \multicolumn{1}{c|}{Parameter} & \multicolumn{1}{c|}{Parameter} & \multicolumn{1}{c|}{Parameter} & \multicolumn{1}{c|}{Switch} & \multicolumn{1}{c|}{Switch} & \multicolumn{1}{c|}{RMSE} & \multicolumn{1}{c|}{RMSE}\\
        \rowcolor{Gray}
        \multirow{-3}{*}{Test}& \multicolumn{1}{c|}{($r$)} & \multicolumn{1}{c|}{($q_x = q_p$)} & \multicolumn{1}{c|}{($\delta$)} & \multicolumn{1}{c|}{($A$)} & \multicolumn{1}{c|}{($B$)} & \multicolumn{1}{c|}{($C$)} & \multicolumn{1}{c|}{($t_0$), hours} & \multicolumn{1}{c|}{($t_s$), hours} & \multicolumn{1}{c|}{} & \multicolumn{1}{c|}{}\\
        \arrayrulecolor{white}\hline
        \rowcolor{LightRed}
        1	& 1E-03	& 1E-04	& 1	& 0.1	& 0	& 0.01	& 2	& 4.99	& 4.9E-03	& 1E-02\\
        \arrayrulecolor{white}\hline
        \rowcolor{LightRed}
        2 & 1E-06	& 1E-04		& 1	& 	0.2		& 0	& 	0		& 0.1	& 	4.93		& 7.8E-03	& 	1.3E-02 \\
        \arrayrulecolor{white}\hline
        \rowcolor{LightGreen}
        3		& 1E-06		& 1E-04	& 	1	& 	0.2	& 	0	& 	0	& 	2	& 	2.01	& 	2.1E-03	& 	3.6E-03 \\
        \arrayrulecolor{white}\hline
        \rowcolor{LightRed}
        4	& 1E-03	& 1E-06	& 1	& 0	& 0	& 0.01	& 2	& 3.04	& 2.4E-04	& 4.8E-04 \\
        \arrayrulecolor{white}\hline
        \rowcolor{LightRed}
        5	& 1E-03	& 1E-06	& 1	& 0.1	& 0	& 0	& 0	& 4.99	& 3E-03	& 4.1E-03 \\
        \arrayrulecolor{white}\hline
        \rowcolor{LightGreen}
        6	& 1E-03	& 1E-06	& 1	& 0.1	& 0	& 0.01	& 2	& 2.01	& 2.6E-04	& 3.4E-04 \\
        \arrayrulecolor{white}\hline
        \rowcolor{LightYellow}
        7	& 1E-03	& 1E-06	& 5	& 0.1	& 0	& 0.01	& 2	& 2.04	& 2.2E-04	& 2.6E-04\\ 
        \arrayrulecolor{white}\hline
        \rowcolor{LightGreen}
        8	& 1E-06	& 1E-06	& 1	& 0	& 0	& 0	& n/a	& 4.99	& 2.6E-04	& 6.4E-04 \\
        \arrayrulecolor{white}\hline
        \rowcolor{LightGreen}
        9	& 1E-06	& 1E-06	& 1	& 0.1	& 0	& 0	& 0	& 0.01	& 2.9E-04	& 2.5E-04\\
        \hline
    \end{tabular}}
\end{table*}

\begin{figure}[ht!]
    \centering
    \begin{subfigure}[b]{0.49\textwidth}
        \centering
        \includegraphics[width=\textwidth]{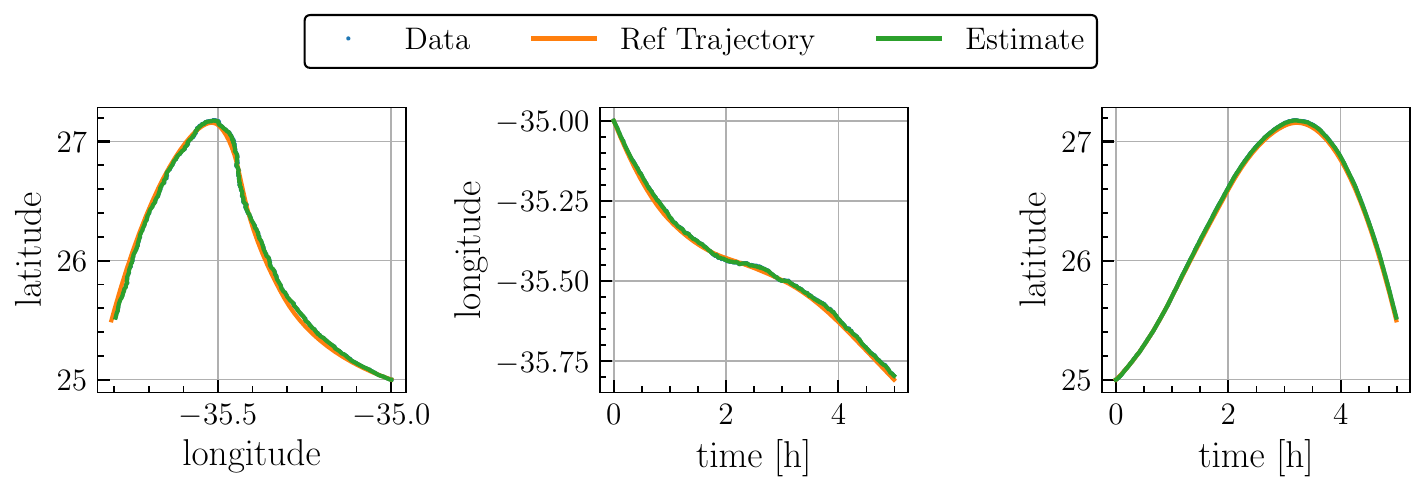}
        \caption{State trajectories}
        \label{fig:balloon_test_8_a}
    \end{subfigure}
    \vfill
    \begin{subfigure}[b]{0.49\textwidth}
        \centering
        \includegraphics[width=\textwidth]{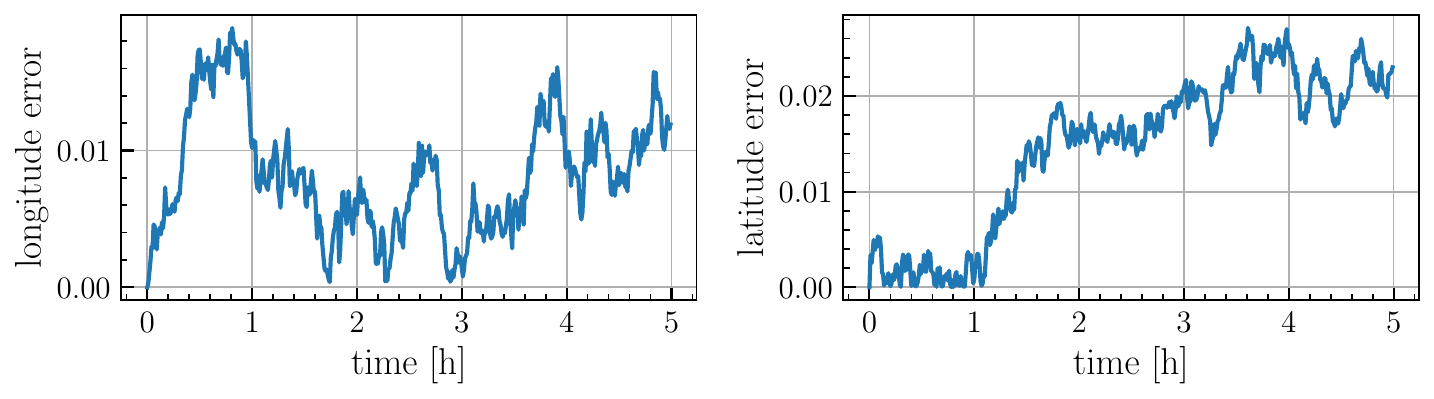}
        \caption{Estimation errors}
        \label{fig:balloon_test_8_b}
    \end{subfigure}
    \vfill
    \begin{subfigure}[b]{0.49\textwidth}
        \centering
        \includegraphics[width=\textwidth]{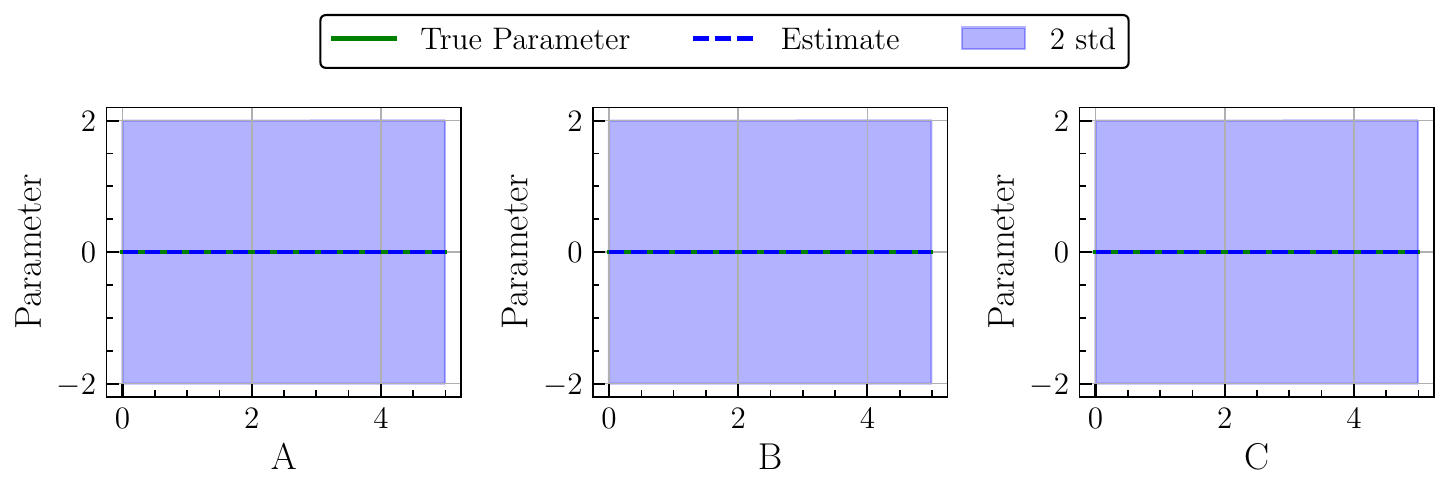}
        \caption{Estimated parameters}
        \label{fig:balloon_test_8_c}
    \end{subfigure}
    \caption{Balloon Test 8: Unbiased simulation with low measurement and process noises. The method performs traditional Gaussian filtering, and accurately recovers the true trajectory of the balloon.}
    \label{fig:balloon_test_8}
\end{figure}

First we describe the unbiased case, where the observation data is not corrupted. The simulated trajectory is shown in the Figure \ref{fig:balloon_test_8_a}. We see that the estimate matches the reference trajectory almost exactly. However, to be more precise we plot the the absolute errors of longitude and latitude in Figure \ref{fig:balloon_test_8_b}. Furthermore, Figure \ref{fig:balloon_test_8_c} we demonstrates that the the parameters have zero mean and unitary standard deviation. This behavior is explained by two facts, the first is that the problem correctly identifies the nominal branch as the best branch (no switching), and second is that {\it no parameter learning of the sensor model} occurs for the nominal branch --- as a result, the prior distribution on the sensor model parameter does not change (rmeains zero mean with identity variance).

\begin{figure}[ht!]
    \centering
    \begin{subfigure}[b]{0.49\textwidth}
        \centering
        \includegraphics[width=\textwidth]{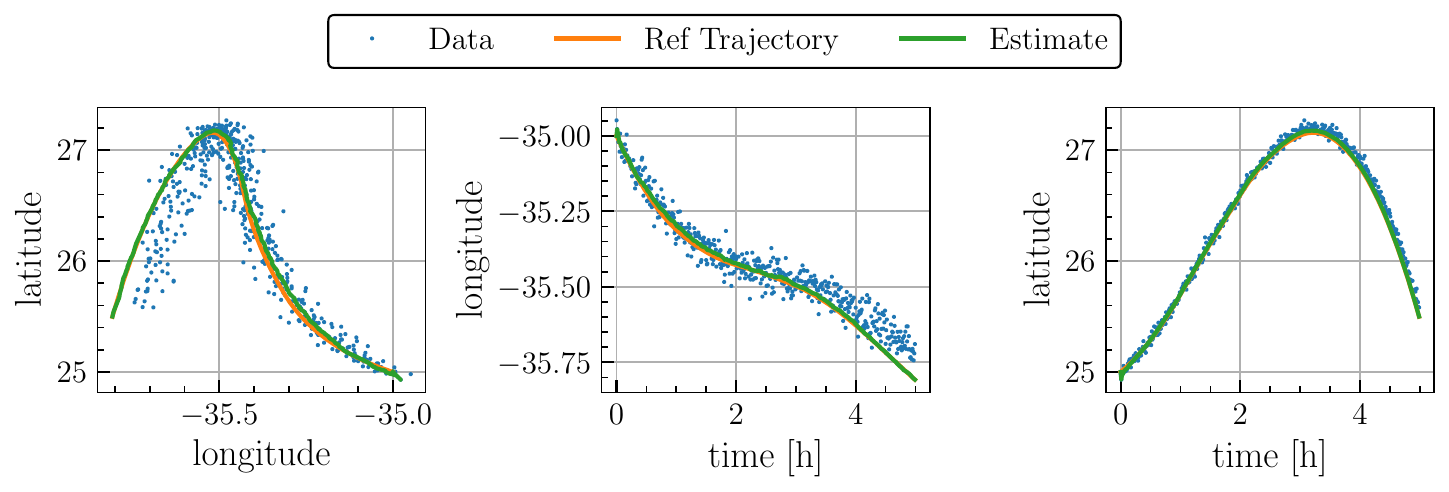}
        \caption{State trajectories}
    \end{subfigure}
    \vfill
    \begin{subfigure}[b]{0.49\textwidth}
        \centering
        \includegraphics[width=\textwidth]{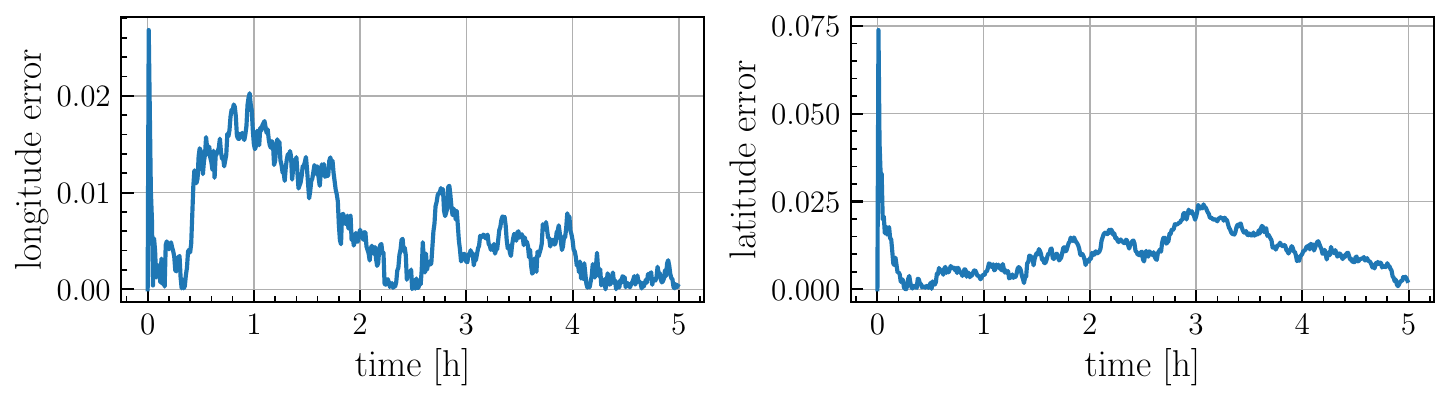}
        \caption{Estimation errors}
    \end{subfigure}
    \vfill
    \begin{subfigure}[b]{0.49\textwidth}
        \centering
        \includegraphics[width=\textwidth]{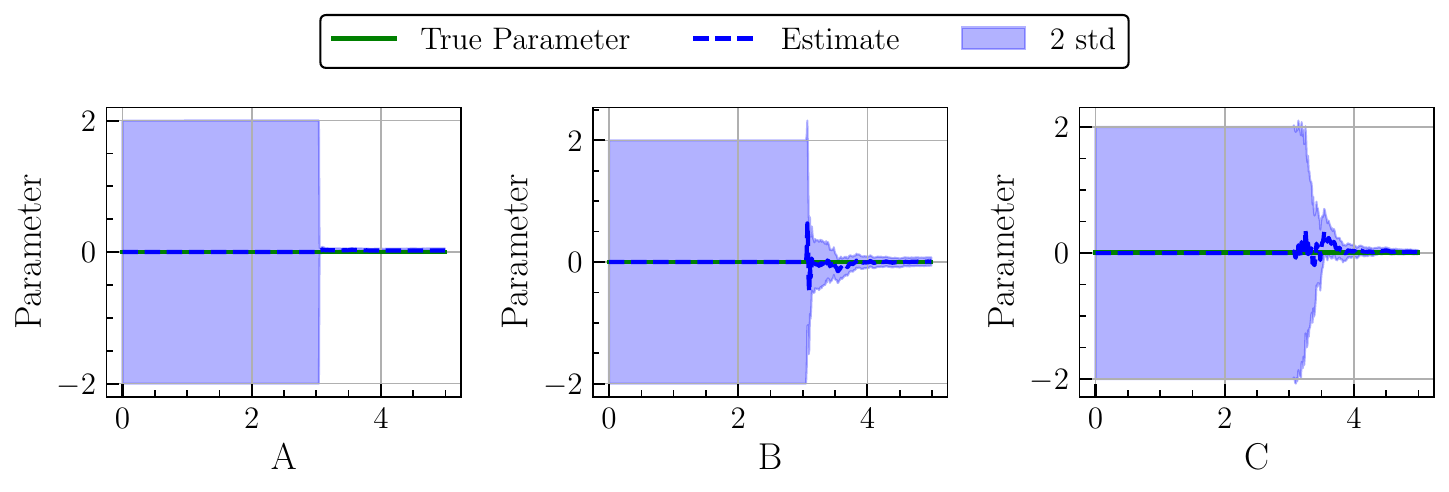}
        \caption{Estimated parameters}
    \end{subfigure}
    \caption{Balloon Test 4: small corruption parameters with large measurement noise and small process noise. True switching time is 2 seconds. Due to small corruption parameters, the SKF does not capture the exact switching time. Due to small bias, it accurately recovers the state. Once the SKF identifies the switching time, the standard deviation of bias parameters significantly decreases because it starts learning the parameters.}
    \label{fig:balloon_test_4}
\end{figure}

\begin{figure}[ht!]
    \centering
    \begin{subfigure}[b]{0.49\textwidth}
        \centering
        \includegraphics[width=\textwidth]{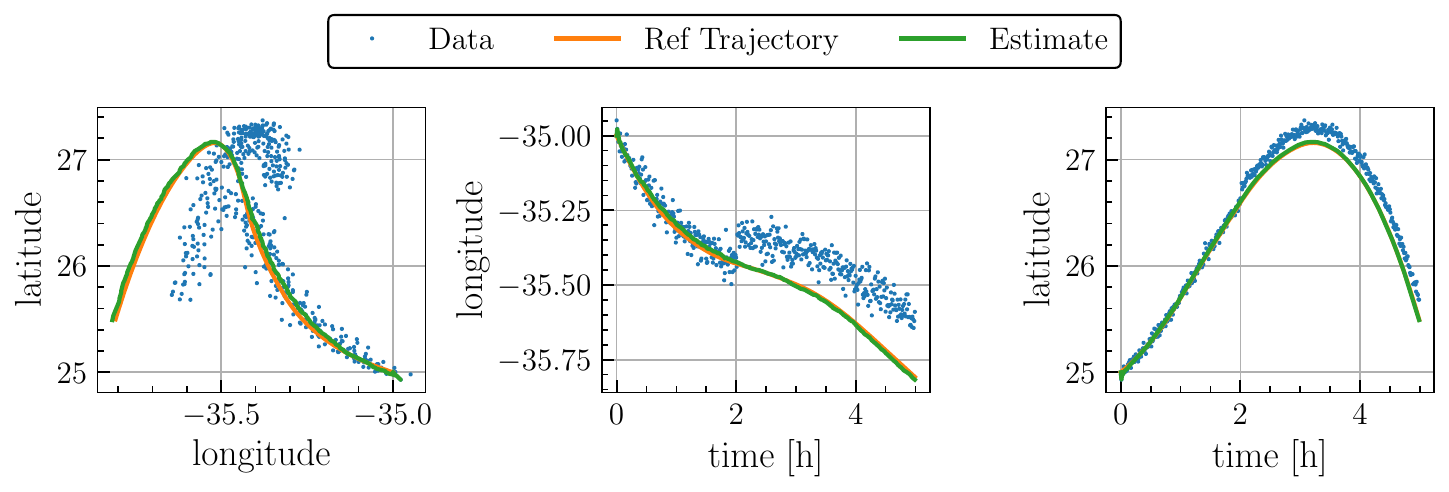}
        \caption{State trajectories}
    \end{subfigure}
    \vfill
    \begin{subfigure}[b]{0.49\textwidth}
        \centering
        \includegraphics[width=\textwidth]{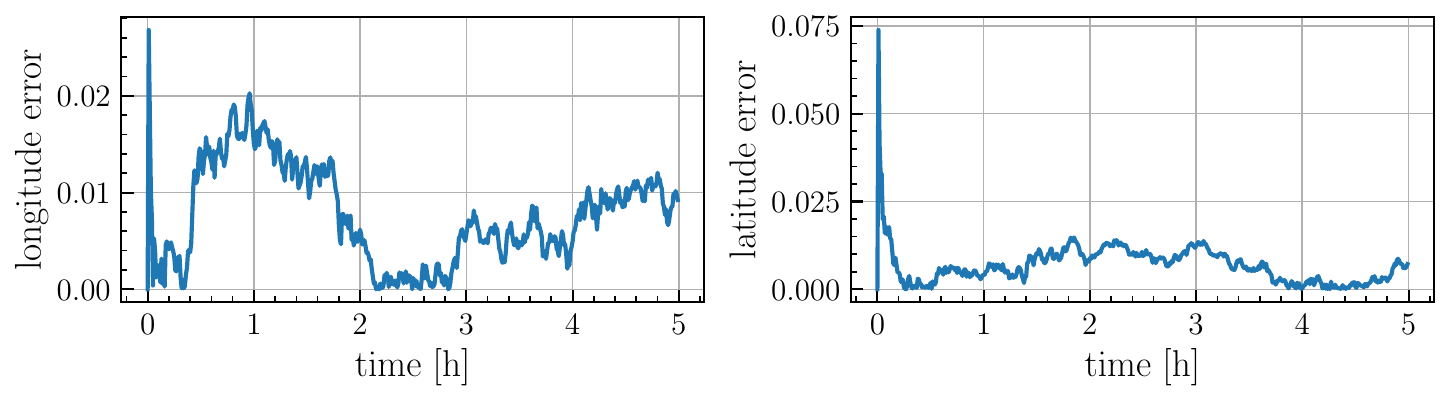}
        \caption{Estimation errors}
    \end{subfigure}
    \vfill
    \begin{subfigure}[b]{0.49\textwidth}
        \centering
        \includegraphics[width=\textwidth]{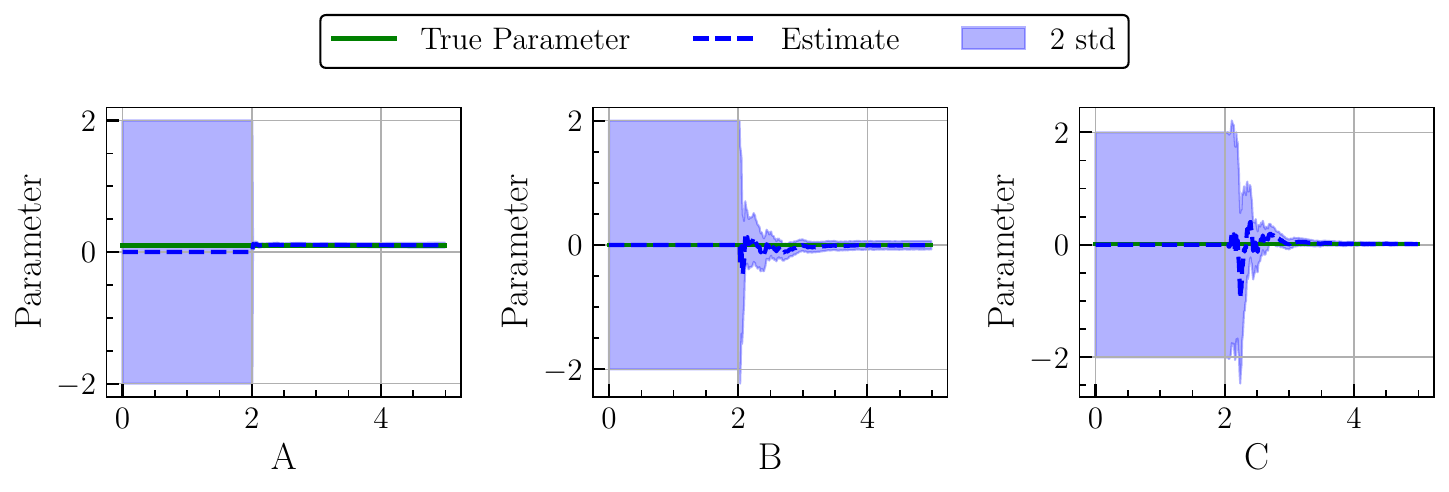}
        \caption{Estimated parameters}
    \end{subfigure}
    \caption{Balloon Test 6: Large corruption parameters with large measurement noise and small process noise. The true switching time is 2 seconds. The system identifies switching time exactly, and regardless of large bias, accurately estimates true states. Once the SKF identifies the switching time, the standard deviation of bias parameters significantly decreases because it starts learning the parameters.}
    \label{fig:balloon_test_6}
\end{figure}

Next, we use Tests 4 and 6 to quantify the effect of parameter magnitude on the performance of the SKF. The simulated results for each test are plotted in Figures \ref{fig:balloon_test_4} and \ref{fig:balloon_test_6} respectively. We see that when applied to Test 4, the SKF is unable to identify the switching time exactly, and has a delay of approximately 100 time steps. As the parameters increase in Test 6, the SKF is able to recover the switching time within a single time step. These results suggest that it is harder to identify the switching time with smaller parameters. In the parameter plots in Figures \ref{fig:balloon_test_4} and \ref{fig:balloon_test_6} we see two clear modes of operation: initial high variance mode, and low variance mode once the corruption is identified. As discussed previously, the SKF does not estimate the parameters until the corruption is identified. Once the SKF identifies the presence of corruption, it begins to update the parameter distribution.

Tests 6 and 7, are used to analyze the impact of sampling frequency on estimator performance. Increasing the time between observations, i.e. decreasing frequency, increases the error in the estimated switching time slightly. However, whether observing once every time step or every 5 time-steps the error was the time between observations. These results suggest that decreasing the sampling frequency decreases the ability of the filter to estimate switching time but not drastically.

Next, Tests 5 and 9 quantify the effect of measurement noise. These tests use identical corruption parameters, sampling frequency, switching time, and process noise. However, the SKF is able to more accurately capture the switching time when applied the Test 9 which has smaller measurement noise. These experiments suggest that smaller measurement noise increases the performance of SKF. To better visualize the magnitude of the measurement noise relative to the overall trajectory, we use table \ref{tab:meas_conv} which list two standard deviations of measurement noise used as the percentage of trajectory range. 

\begin{table}[ht!]
    \centering
    \caption{Measurement noise values as a fraction of trajectory range.}
    \begin{tabular}{|c|c|c|c|c|c|}
    \hline
    Measurement noise value & 1e-6 & 1e-5 & 5e-5 & 1e-4 & 1e-3\\
    \hline
    $2\sigma$ to range percent ratio & 0.25 & 0.78 & 1.75 & 2.47 & 7.81 \\
    \hline
    \end{tabular}
    \label{tab:meas_conv}
\end{table}

Next, we analyze Tests 2 and 3 to quantify the impact of the initial onset of the corruption on the performance of the SKF. In Test 2 we initiate the corruption near the start of the simulation, while in Test 3, it activates midway through. The results indicate that the filter struggles to precisely identify the early switching time but is able to detect the later switching time with the accuracy of one time step. This discrepancy could be attributed to the filter's initial large covariance, which introduces greater uncertainty in the early stages of the simulation. As the filter converges and gains confidence, it becomes better equipped to detect changes in system behavior, and identify the start of the corruption. 

Finally, we examine the effect of process noise by comparing Tests 1 and 6. Test 1 has higher process noise but is unable to recover the switching time, whereas Test 6, with lower process noise, is able to identify the exact switching time within a single time step. This observation suggests that higher process noise may hinder the ability to accurately identify switching times, while lower process noise appears to improve the filter's performance in this regard.

\subsubsection{Statistical Analysis}\label{sec:statistical_balloon}

In this subsection, we describe the results of a statistical analysis that we use to further examine the effects of various problem setup parameters on the method's ability to accurately estimate the switching time and reference balloon trajectory. Specifically, we conduct 2500 different experiments by varying parameter process noise, measurement noise, and the corruption parameters. The range of values assigned to each of these variables, is listed in Table \ref{tab:test_values_balloon_sa}. All combinations of these variables are used in the test set and the state process noise is a fixed function of the measurement noise

\begin{table}[ht!]
    \centering
    \caption{Values of test setup variables for the statistical analysis of the drifting balloon problem.}
    \begin{tabular}{|c|c|}
    \hline
    Variable & Tested Values \\
    \hline
    Parameter process noise ($q_p$) & [1e-10, 1e-8, 1e-6, 1e-4, 1e-2] \\
    Measurement Noise ($r$) & [1e-6, 1e-5, 5e-5, 1e-4] \\
    Static Bias Parameter ($A$) & [0, 0.001, 0.01, 0.1, 0.5] \\
    Linear Bias Parameter ($B$)& [0, 0.001, 0.01, 0.1, 2.0] \\
    Quadratic Bias Parameter ($C$)& [0, 0.001, 0.01, 0.1, 2.0] \\
    \hline
    \end{tabular}
    \label{tab:test_values_balloon_sa}
\end{table}

In the following we summarize the performance of the SKF corresponding to each of these variables in figure pairs: the first figure representing the percentage of successful simulations, where all but the chosen variable is varied; and the second figure of a pair plots the median RMSE obtained over all these other simulations.

\begin{figure}[htbp]
    \centering
        \includegraphics[width=0.49\textwidth]{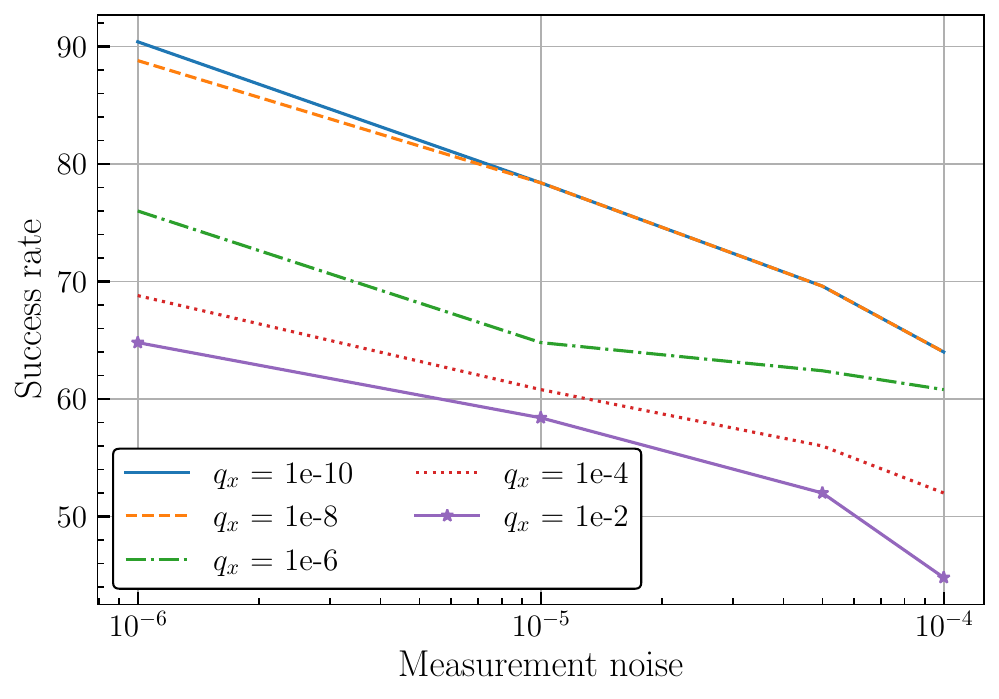}
        \caption{Successful simulations as a function of measurement noise for fixed values of parameter process noise. As the process and measurement noise increases, it is harder to identify switching time.}
    \label{fig:sa_measurement}
\end{figure}

\begin{figure}[htbp]
    \centering
        \includegraphics[width=0.49\textwidth]{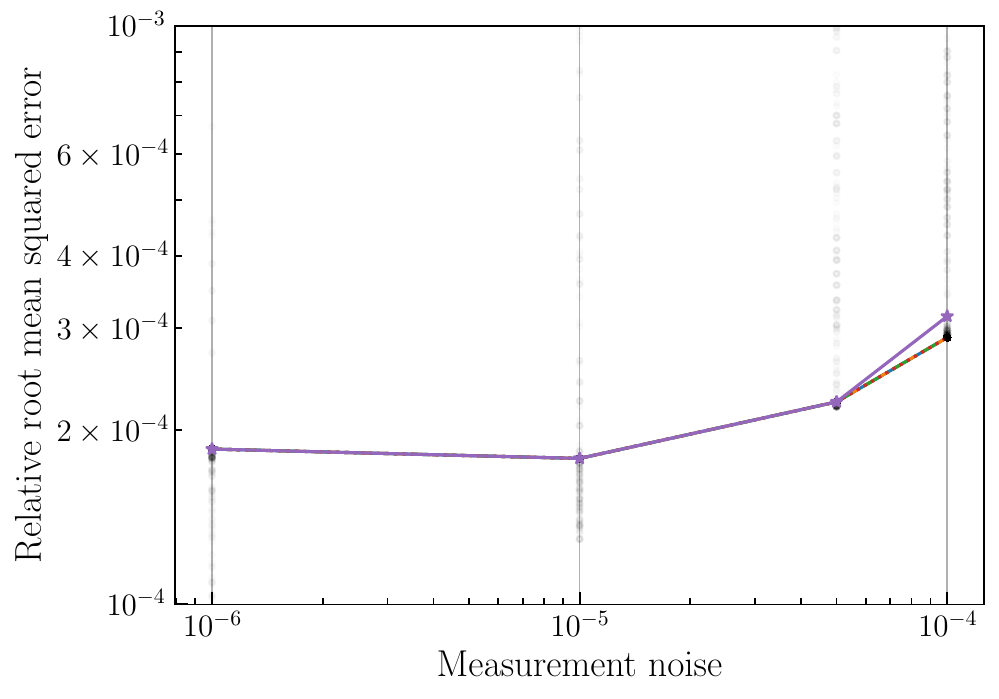}
        \caption{Estimated trajectory $RMSE$ as a function of measurement noise for fixed values of parameter process noise. Each black dot represents the $RMSE$ of a performed simulation for a given value of measurement noise. An increase in measurement noise correlates to an increase in $RMSE$. }
    \label{fig:sa_measurement_rmse}
\end{figure}

Figures \ref{fig:sa_measurement} and \ref{fig:sa_measurement_rmse} show that reducing the measurement noise improves the method’s ability to accurately estimate the switching time. We also note that as the measurement noise increases, the $RMSE$ of the estimate trajectory also increases. In contrast, Figures 11 - 16 quantify the impact of the bias parameters on performance. Each of the parameters (static, linear, and quadratic) is simulated as an individual variable. Specifically, Figures 11,13, and 15, highlight the method’s ability to capture switching time increasing with increasing bias. Indeed, for large (static and linear) bias parameters, the true switching time can be identified within a single time step for every performed simulation. Intuitively, this results makes sense as it should be easier to identify a corruption when an abrupt change in the observational data is observed. Additionally, Figures 12, 14, and 16 show that the estimated trajectory $RMSE$ is  small for both small and large parameters, but increases for values between the two extremes. For smaller values of the corruption parameters, the trajectory estimation errors are small because the corruption introduced to the measurements is minimal and does not significantly perturb the estimate of the state. Additionally, at large parameter values, it becomes easier to identify the presence of corruption and estimate the parameters of the bias model using the state augmentation approach. Therefore, it also becomes easy to explicitly negate the effects of this corruption. 

\begin{figure}[ht!]
    \centering
        \includegraphics[width=0.49\textwidth]{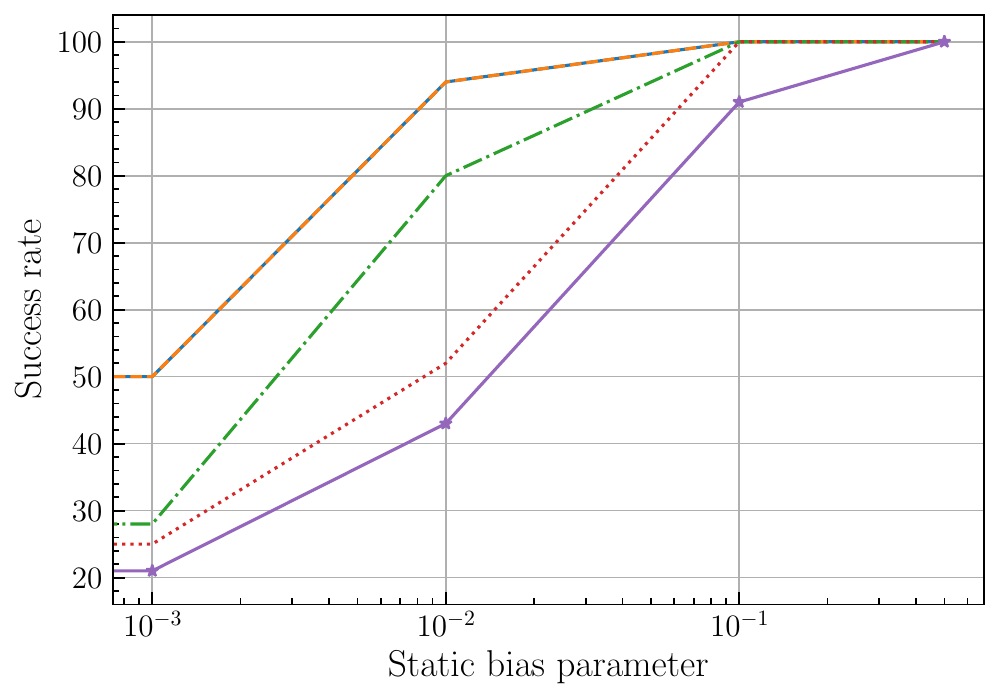}
        \caption{Successful simulations as a function of static bias parameter for fixed values of parameter process noise. Larger static parameters correspond to an increased success rate of identifying switching time.}
    \label{fig:sa_static}
\end{figure}

\begin{figure}[ht!]
    \centering
        \includegraphics[width=0.49\textwidth]{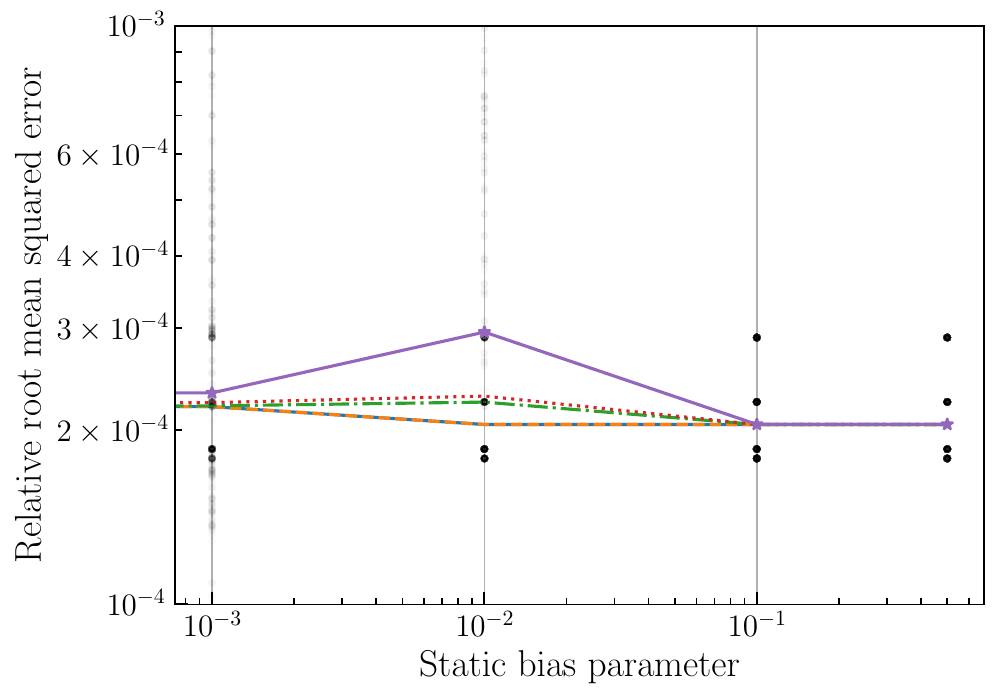}
        \caption{Estimated trajectory $RMSE$ as a function of static bias parameter for fixed values of parameter process noise. Each black dot represents the $RMSE$ of a performed simulation for a given value of static bias parameter.}
    \label{fig:sa_static_rmse}
\end{figure}

\begin{figure}[ht!]
    \centering
        \includegraphics[width=0.49\textwidth]{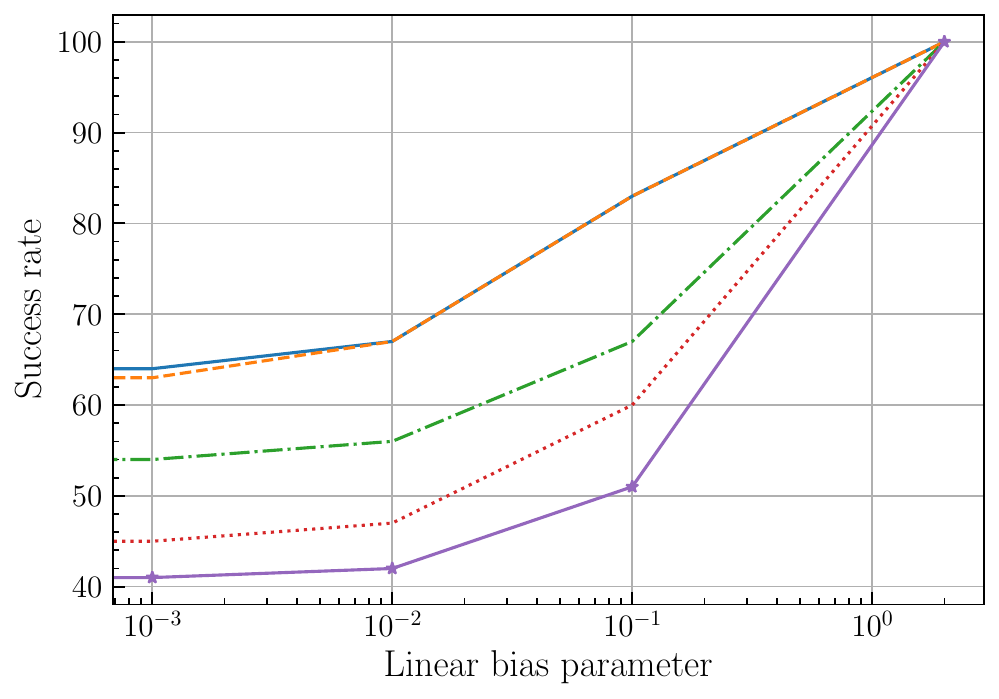}
        \caption{Successful simulations as a function of linear bias parameter for fixed values of parameter process noise. Larger linear parameters correspond to an increased success rate of identifying switching time.}
    \label{fig:sa_linear}
\end{figure}

\begin{figure}[ht!]
    \centering
        \includegraphics[width=0.49\textwidth]{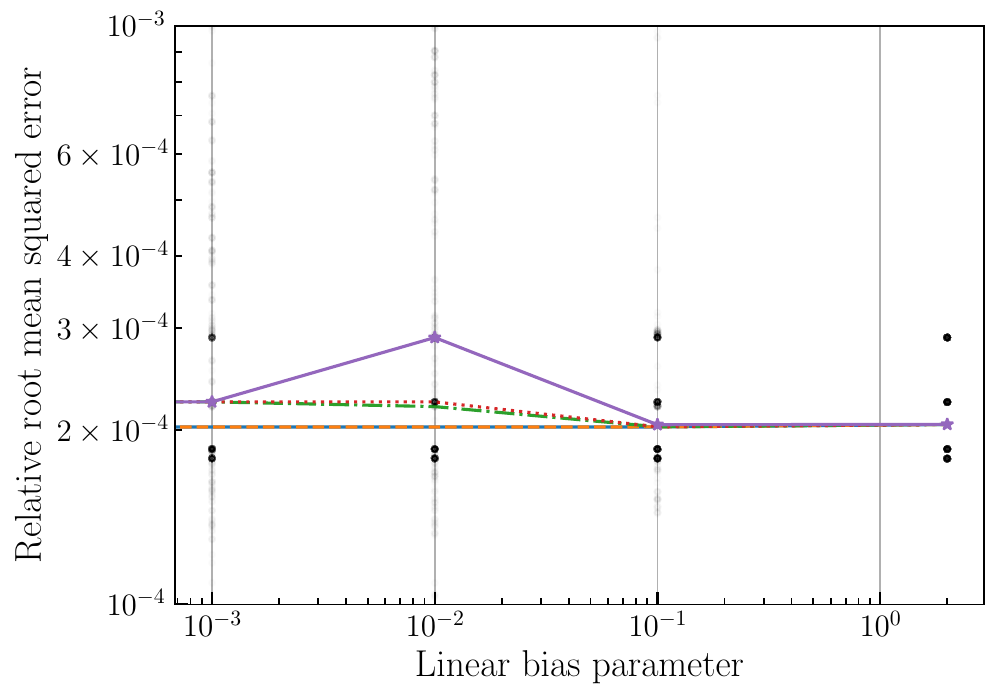}
        \caption{Estimated trajectory $RMSE$ as a function of linear bias parameter for fixed values of parameter process noise. Each black dot represents the $RMSE$ of a performed simulation for a given value of linear bias parameter.}
    \label{fig:sa_linear_rmse}
\end{figure}

\begin{figure}[ht!]
    \centering
        \includegraphics[width=0.49\textwidth]{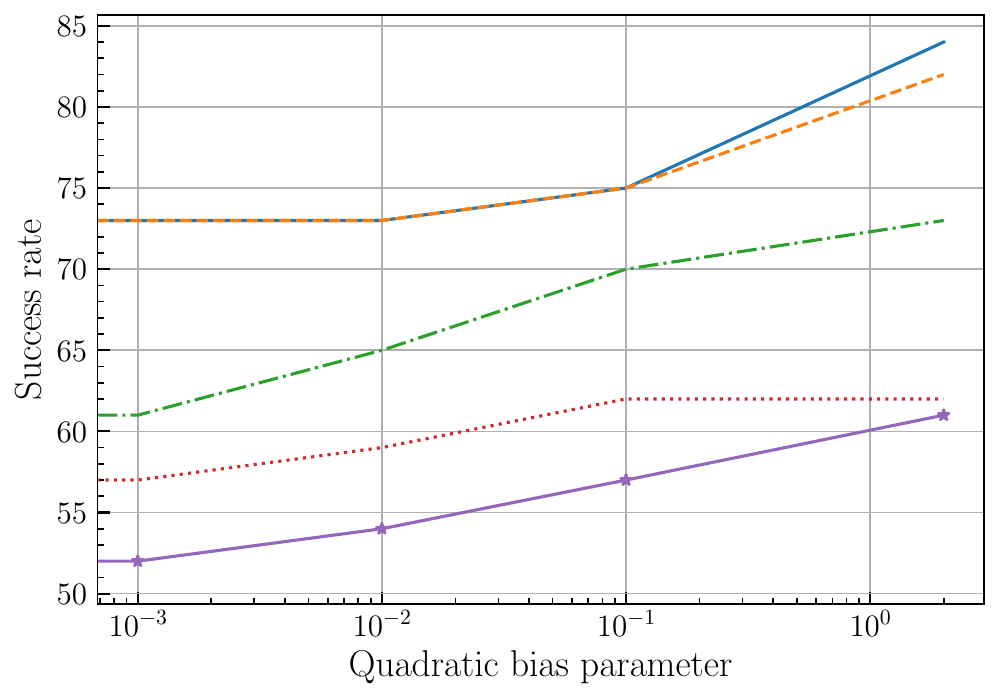}
        \caption{Successful simulations as a function of quadratic bias parameter for fixed values of parameter process noise. Larger quadratic parameters correspond to an increased success rate of identifying switching time.}
    \label{fig:sa_quadratic}
\end{figure}

\begin{figure}[ht!]
    \centering
        \includegraphics[width=0.49\textwidth]{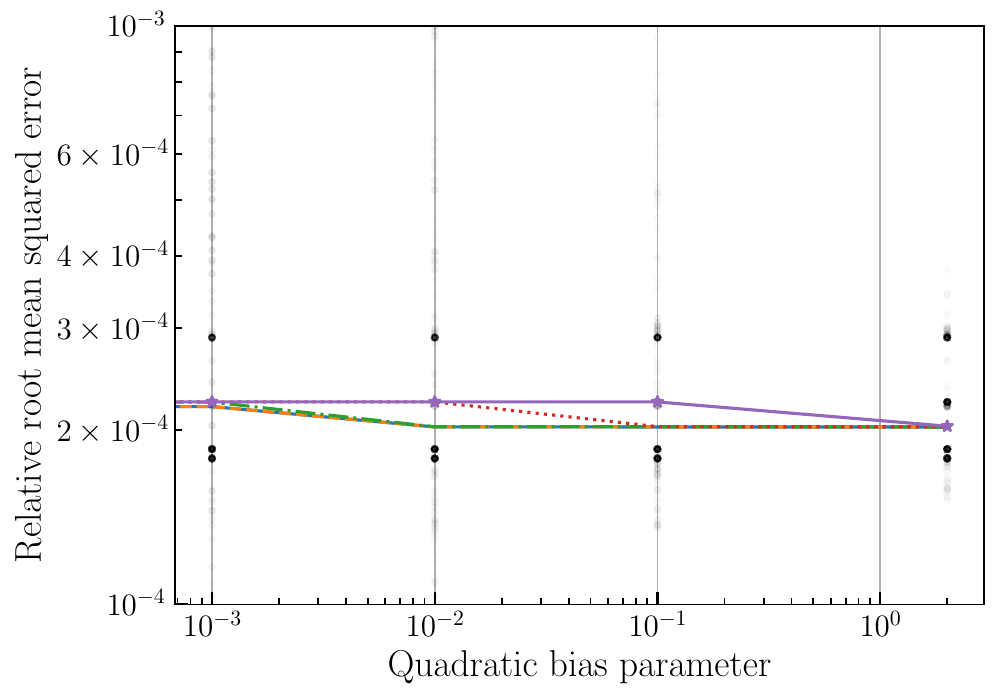}
        \caption{Estimated trajectory $RMSE$ as a function of quadratic bias parameter for fixed values of parameter process noise. Each black dot represents the $RMSE$ of a performed simulation for a given value of quadratic bias parameter.}
    \label{fig:sa_quadratic_rmse}
\end{figure}

Overall, we see that in cases where the bias is large, we are able to reliably identify the bias and correct for its errors. When the bias is small, the state estimation error is also small. It is possible that there is a middle regime where the bias is just large enough to cause an error, but to remain undetected and uncorrected.

\subsection{Shuttle Reentry}

\subsubsection{Problem setup}

\begin{figure}[ht!]
    \centering
        \includegraphics[width=0.49\textwidth]{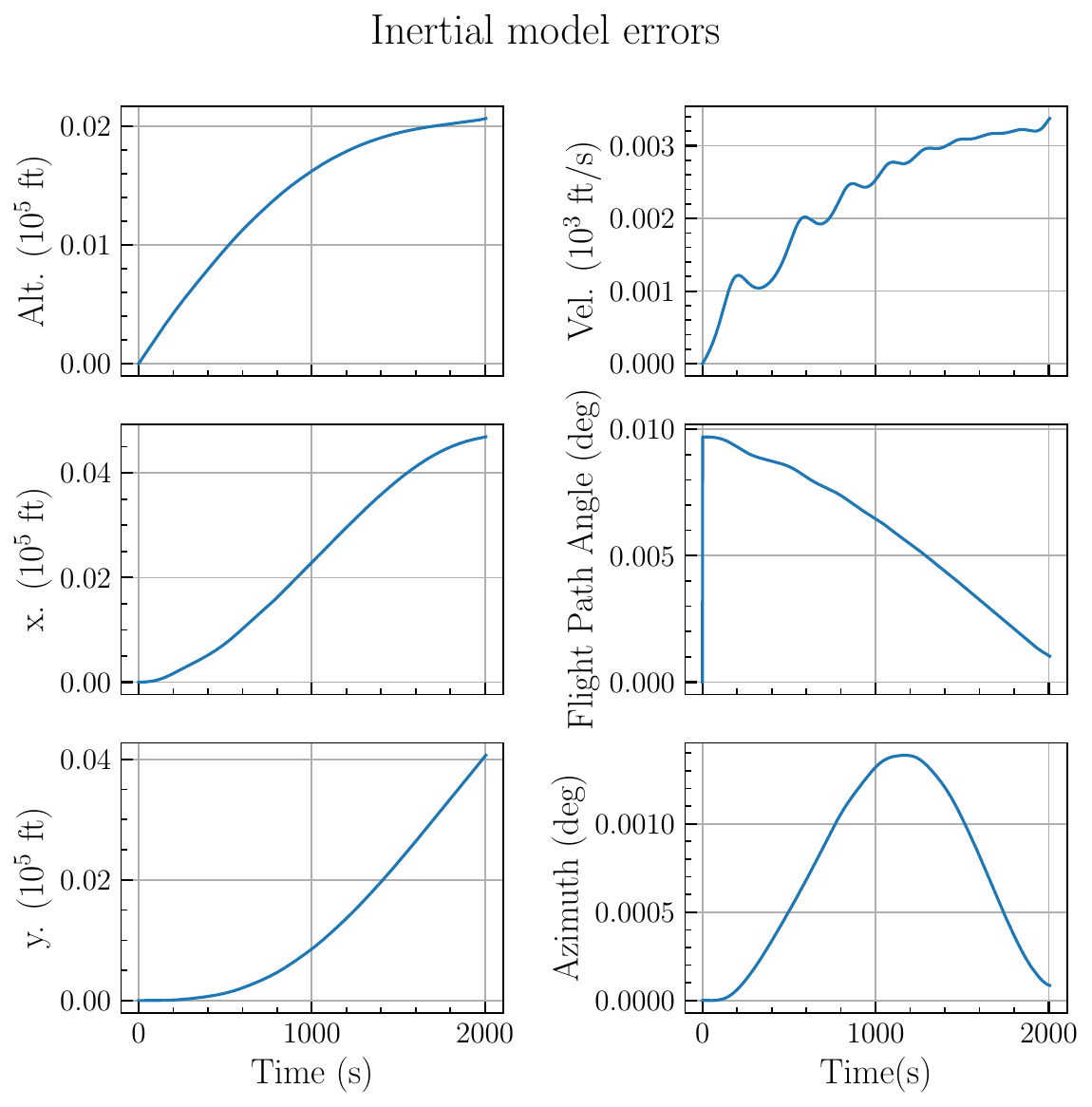}
        \caption{Absolute errors between the inertial and non-inertial model readings. There is a significant positional and velocity difference that grows over time, the attitude states do not grow over time.}
    \label{fig:IM_error}
\end{figure}

Next we consider a more realistic inertial navigation problem simulating a space shuttle re-entry. The reference simulation is the same as described in \cite[Chap 6.1]{betts2010practical}. However, we corrupt the reference trajectory to provide GPS measurements to the estimation system by adding random noise and corruption biases. In particular, we extract the forces from the reference simulation to generate the reference measurements using a zero-order hold for an inertial model that evolves with a time step of 1.4 seconds. The absolute error difference between inertial and (noiseless) non inertial models is demonstrated the Figure \ref{fig:IM_error}. All the relative errors in subsequent results are with respect to this noiseless inertial model trajectory. The simulations have  a switching time of 500 seconds and initial position and parameter variances of 0.001 and 1, respectively. Each conducted experiment setup is documented in the Table \ref{tab:shuttle_results}.

Specifically, we analyze the effects of different parameters, measurement noise, and process noise on the performance of the SKF. For this set of experiments, we only observe the positions using the GPS --- the other states were considered unobserved by external sensor. Moreover, we also introduce a cap to the corruption bias model --- once the corruption reaches 1000 meter offset, it remains at that value for the rest of the simulation, that is, the true corruption model is 

\begin{equation}
    H_b = 
    \begin{cases}
        A+B(t-t_s)+C(t-t_s)^2, \quad H_b<1000 \\
        1000,\quad otherwise
    \end{cases}
    \label{eq:capped_quad}
\end{equation}
Even though the true corruption model is capped, the model we are learning is still quadratic. As a result, we cannot guarantee that the correct bias parameters are learned. Nevertheless, we often still identify the correct switching time and maintain low state errors. The experimental parameters and results are presented in Table \ref{tab:shuttle_results}. Each test is color coded in the same way as in the drifting balloon example above.

\begin{table*}[htbp]
    \centering
    \caption{Shuttle Reentry Setup and Results}
    \label{tab:shuttle_results}
    \resizebox{\textwidth}{!}{
    \begin{tabular}{|c|ccccccccccccc|}
        \arrayrulecolor{white}\hline
        \rowcolor{Gray}
        & \multicolumn{1}{c|}{Static} & \multicolumn{1}{c|}{Linear} & \multicolumn{1}{c|}{Quad.} & \multicolumn{1}{c|}{Process} & \multicolumn{1}{c|}{Process} & \multicolumn{1}{c|}{Meas.} & \multicolumn{1}{c|}{Estimated} & \multicolumn{1}{c|}{Altitude} & \multicolumn{1}{c|}{Longitude} & \multicolumn{1}{c|}{Latitude} & \multicolumn{1}{c|}{Speed} &\multicolumn{1}{c|}{Flight Path} &\multicolumn{1}{c|}{Azimuth}\\
        \rowcolor{Gray}
        \multirow{-2}{*}{Test}& \multicolumn{1}{c|}{Param.} & \multicolumn{1}{c|}{Param.} & \multicolumn{1}{c|}{Param.} & \multicolumn{1}{c|}{Noise x} & \multicolumn{1}{c|}{Noise $\theta$} & \multicolumn{1}{c|}{Noise} & \multicolumn{1}{c|}{Switch} & \multicolumn{1}{c|}{RMSE} & \multicolumn{1}{c|}{RMSE} & \multicolumn{1}{c|}{RMSE}& \multicolumn{1}{c|}{RMSE}& \multicolumn{1}{c|}{RMSE}& \multicolumn{1}{c|}{RMSE}\\
        \hline
        \rowcolor{LightGreen}
        1&	0&	0&	0&	1E-08&	1E-12&	1E-08&	1970.1&	2.3E-04	&	2.3E-06	&	2.1E-05	&	1.7E-03	&	5.9E-02	&	3.6E-03\\
        \arrayrulecolor{white}\hline
        \rowcolor{LightRed}
        2 &	0	&	0	&	1	&	1E-08	&	1E-12	&	1E-08	&	1968.7	&	1.6E-02	&	1.4E-04	&	3.9E-04	&	1.5E-03	&	6.9E-02	&	1.5E-02\\
        \arrayrulecolor{white}\hline
        \rowcolor{LightYellow}
        3	&	0	&	0	&	10	&	1E-08	&	1E-12	&	1E-08	&	513.5	&	3.2E+00	&	2.4E-02	&	1.2E-01	&	7.4E-02	&	5.6E+00	&	2.0E-01\\
        \arrayrulecolor{white}\hline
        \rowcolor{LightRed}
        4	&	0	&	1	&	0	&	1E-08	&	1E-12	&	1E-08	&	1971.5	&	1.2E-02	&	1.1E-04	&	3.1E-04	&	1.7E-03	&	5.7E-02	&	3.2E-03\\
        \arrayrulecolor{white}\hline
        \rowcolor{LightGreen}
        5	&	0	&	1	&	1	&	1E-08	&	1E-12	&	1E-08	&	500.9	&	9.0E-05	&	1.7E-06	&	7.8E-06	&	2.2E-03	&	5.2E-02	&	8.3E-03\\
        \arrayrulecolor{white}\hline
        \rowcolor{LightYellow}
        6	&	0	&	1	&	10	&	1E-08	&	1E-12	&	1E-08	&	513.5	&	3.1E+00	&	2.1E-02	&	1.2E-01	&	7.4E-02	&	5.5E+00	&	1.2E-01\\
        \arrayrulecolor{white}\hline
        \rowcolor{LightRed}
        7	&	0	&	10	&	0	&	1E-08	&	1E-12	&	1E-08	&	1970.1	&	1.6E-02	&	1.4E-04	&	4.0E-04	&	1.7E-03	&	6.1E-02	&	4.0E-03\\
        \arrayrulecolor{white}\hline
        \rowcolor{LightYellow}
        8	&	0	&	10	&	1	&	1E-08	&	1E-12	&	1E-08	&	498.1	&	6.1E-05	&	4.2E-06	&	1.9E-05	&	2.2E-03	&	4.7E-02	&	6.2E-03\\
        \arrayrulecolor{white}\hline
        \rowcolor{LightYellow}
        9	&	0	&	10	&	10	&	1E-08	&	1E-12	&	1E-08	&	512.1	&	3.1E+00	&	1.9E-02	&	1.4E-01	&	7.4E-02	&	5.5E+00	&	1.7E-01\\
        \arrayrulecolor{white}\hline
        \rowcolor{LightGreen}
        10	&	0	&	100	&	0	&	1E-08	&	1E-12	&	1E-08	&	499.5	&	9.3E-06	&	3.7E-06	&	1.5E-05	&	2.1E-03	&	4.3E-02	&	1.1E-02\\
        \arrayrulecolor{white}\hline
        \rowcolor{LightGreen}
        11	&	0	&	100	&	1	&	1E-08	&	1E-12	&	1E-08	&	499.5	&	4.5E-06	&	1.7E-06	&	1.0E-05	&	2.4E-03	&	5.1E-02	&	5.3E-03\\
        \arrayrulecolor{white}\hline
        \rowcolor{LightGreen}
        12	&	0	&	100	&	10	&	1E-08	&	1E-12	&	1E-08	&	499.5	&	2.7E-05	&	2.6E-06	&	1.4E-05	&	2.0E-03	&	5.3E-02	&	1.1E-02\\
        \arrayrulecolor{white}\hline
        \rowcolor{LightGreen}
        13	&	100	&	0	&	0	&	1E-08	&	1E-12	&	1E-08	&	499.5	&	1.4E-02	&	1.2E-04	&	3.7E-04	&	1.8E-03	&	5.1E-02	&	9.0E-03\\
        \arrayrulecolor{white}\hline
        \rowcolor{LightGreen}
        14	&	100	&	0	&	1	&	1E-08	&	1E-12	&	1E-08	&	499.5	&	1.3E-04	&	4.6E-06	&	4.6E-05	&	3.7E-03	&	1.1E-01	&	1.4E-02\\
        \arrayrulecolor{white}\hline
        \rowcolor{LightYellow}
        15	&	100	&	0	&	10	&	1E-08	&	1E-12	&	1E-08	&	498.1	&	1.1E-04	&	2.0E-06	&	2.0E-05	&	2.0E-03	&	5.4E-02	&	1.0E-02\\
        \arrayrulecolor{white}\hline
        \rowcolor{LightGreen}
        16	&	100	&	1	&	0	&	1E-08	&	1E-12	&	1E-08	&	499.5	&	5.7E-03	&	4.8E-05	&	1.5E-04	&	1.8E-03	&	4.9E-02	&	8.5E-03\\
        \arrayrulecolor{white}\hline
        \rowcolor{LightGreen}
        17	&	100	&	1	&	1	&	1E-08	&	1E-12	&	1E-08	&	499.5	&	1.4E-04	&	5.4E-06	&	4.8E-05	&	3.6E-03	&	1.2E-01	&	1.8E-02\\
        \arrayrulecolor{white}\hline
        \rowcolor{LightYellow}
        18	&	100	&	1	&	10	&	1E-08	&	1E-12	&	1E-08	&	498.1	&	1.6E-04	&	1.8E-05	&	1.4E-04	&	3.7E-03	&	6.0E-02	&	3.8E-02\\
        \arrayrulecolor{white}\hline
        \rowcolor{LightRed}
        19	&	100	&	10	&	0	&	1E-08	&	1E-12	&	1E-08	&	1968.7	&	1.6E-02	&	1.4E-04	&	3.5E-04	&	5.5E-03	&	7.1E-02	&	2.8E-02\\
        \arrayrulecolor{white}\hline
        \rowcolor{LightGreen}
        20	&	100	&	10	&	1	&	1E-08	&	1E-12	&	1E-08	&	499.5	&	6.2E-05	&	3.7E-06	&	1.7E-05	&	2.3E-03	&	5.0E-02	&	8.6E-03\\
        \arrayrulecolor{white}\hline
        \rowcolor{LightGreen}
        21	&	100	&	10	&	10	&	1E-08	&	1E-12	&	1E-08	&	499.5	&	7.9E-05	&	2.0E-06	&	1.2E-05	&	1.9E-03	&	5.3E-02	&	9.3E-03\\
        \arrayrulecolor{white}\hline
        \rowcolor{LightGreen}
        22	&	100	&	100	&	0	&	1E-08	&	1E-12	&	1E-08	&	499.5	&	5.9E-06	&	1.1E-06	&	7.8E-06	&	2.2E-03	&	5.1E-02	&	4.3E-03\\
        \arrayrulecolor{white}\hline
        \rowcolor{LightGreen}
        23	&	100	&	100	&	1	&	1E-08	&	1E-12	&	1E-08	&	499.5	&	3.9E-06	&	8.1E-07	&	6.5E-06	&	2.1E-03	&	5.2E-02	&	5.5E-03\\
        \arrayrulecolor{white}\hline
        \rowcolor{LightGreen}
        24	&	100	&	100	&	10	&	1E-08	&	1E-12	&	1E-08	&	499.5	&	2.7E-05	&	2.8E-06	&	1.9E-05	&	2.1E-03	&	5.6E-02	&	1.1E-02\\
        \arrayrulecolor{white}\hline
        \rowcolor{LightGreen}
        25	&	0	&	0	&	0	&	1E-08	&	1E-12	&	1E-09	&	1971.5	&	2.3E-04	&	5.7E-07	&	1.8E-05	&	2.2E-03	&	5.9E-02	&	3.1E-03\\
        \arrayrulecolor{white}\hline
        \rowcolor{LightYellow}
        26	&	0	&	1	&	10	&	1E-08	&	1E-12	&	1E-09	&	513.5	&	3.1E+00	&	2.3E-02	&	1.2E-01	&	7.6E-02	&	5.5E+00	&	1.3E-01\\
        \arrayrulecolor{white}\hline
        \rowcolor{LightGreen}
        27	&	0	&	100	&	10	&	1E-08	&	1E-12	&	1E-09	&	499.5	&	2.8E-05	&	1.0E-06	&	1.1E-05	&	2.3E-03	&	6.7E-02	&	1.4E-02\\
        \arrayrulecolor{white}\hline
        \rowcolor{LightGreen}
        28	&	0	&	0	&	0	&	1E-08	&	1E-12	&	1E-06	&	1968.7	&	2.8E-04	&	1.2E-05	&	1.5E-04	&	1.8E-03	&	5.8E-02	&	7.0E-03\\
        \arrayrulecolor{white}\hline
        \rowcolor{LightYellow}
        29	&	0	&	1	&	10	&	1E-08	&	1E-12	&	1E-06	&	513.5	&	3.0E+00	&	2.4E-02	&	9.9E-02	&	6.1E-02	&	5.3E+00	&	1.1E-01\\
        \arrayrulecolor{white}\hline
        \rowcolor{LightGreen}
        30	&	0	&	100	&	10	&	1E-08	&	1E-12	&	1E-06	&	499.5	&	2.7E-05	&	2.9E-05	&	1.7E-04	&	2.2E-03	&	6.8E-02	&	1.0E-02\\
        \arrayrulecolor{white}\hline
        \rowcolor{LightGreen}
        31	&	0	&	0	&	0	&	1E-06	&	1E-10	&	1E-08	&	2005.1	&	4.0E-06	&	1.2E-07	&	5.9E-07	&	1.5E-03	&	3.3E-02	&	2.0E-02\\
        \arrayrulecolor{white}\hline
        \rowcolor{LightRed}
        32	&	0	&	1	&	10	&	1E-06	&	1E-10	&	1E-08	&	2005.1	&	1.6E-02	&	1.4E-04	&	4.1E-04	&	2.0E-03	&	5.4E-02	&	6.6E-02\\
        \arrayrulecolor{white}\hline
        \rowcolor{LightGreen}
        33	&	0	&	100	&	10	&	1E-06	&	1E-10	&	1E-08	&	499.5	&	6.0E-05	&	4.9E-07	&	8.6E-06	&	4.8E-04	&	3.5E-02	&	1.3E-02\\
        \arrayrulecolor{white}\hline
        \rowcolor{LightGreen}
        34	&	0	&	0	&	0	&	1E-10	&	1E-14	&	1E-08	&	1974.3	&	2.7E-04	&	4.0E-05	&	3.6E-04	&	5.9E-03	&	9.0E-02	&	4.8E-03\\
        \arrayrulecolor{white}\hline
        \rowcolor{LightGreen}
        35	&	0	&	1	&	10	&	1E-10	&	1E-14	&	1E-08	&	500.9	&	4.4E-05	&	4.2E-05	&	2.9E-04	&	9.0E-03	&	1.7E-01	&	8.5E-03\\
        \arrayrulecolor{white}\hline
        \rowcolor{LightGreen}
        36	&	0	&	100	&	10	&	1E-10	&	1E-14	&	1E-08	&	499.5	&	6.2E-06	&	3.6E-05	&	2.8E-04	&	8.7E-03	&	1.7E-01	&	6.7E-03\\
        \arrayrulecolor{white}\hline
    \end{tabular}
    }
\end{table*}

The state of the system, defined in Eq. (\ref{eq:state}), contains parameters with differing units.  To create process and measurement noises that have similar relative errors across states, we multiply the nominal state process noise described in the sections below by the mean absolute values of each state obtained from the reference trajectory. Specifically, when we provide a process noise $q_x$ and a measurement noise $r$ in the below experiments, the actual process and measurement noise of the dynamics are scaled versions $\hat{q}_x$ and $\hat{r}$, respectively, obtained by 
\begin{equation}
    \bar{q_x} = q_x \left[\bar{h}\quad\bar{L}\quad\bar{\lambda}\quad\bar{v}\quad\bar{\gamma}\quad\bar{\alpha} \quad \bar{\Phi}\quad\bar{\Theta}\quad\bar{\Psi}\right]
\end{equation}
\begin{equation}
    \bar{r} = r \left[\bar{h} \quad\bar{L}\quad \bar{\lambda}\right]
\end{equation}
where $\bar{(\cdot)}$ represents the scaling factor for each state. These scaling factors are listed in Table \ref{tab:scaling}. For this problem, only altitude, longitude and latitude are observed, therefore, the measurement noise vector contains only these three states. 
\begin{table*}[ht!]
    \centering
    \caption{Initial conditions and absolute mean values of each state used to scale the process and measurement noises}
    \begin{tabular}{|c|c|c|c|c|c|c|c|c|c|}
    \hline
    State & $\bar{h}$ & $\bar{L}$ & $\bar{\lambda}$ & $\bar{v}$ & $\bar{\gamma}$ & $\bar{\alpha}$ & $\bar{\Phi}$ & $\bar{\Theta}$ & $\bar{\Psi}$ \\
    \hline
    Scaling factor & $1.5e+5$ & $9.3e-1$ & $3.2e-1$ & $1.4e+4$ & $3e-2$ & $8e-1$ & $6e-1$ & $2e-1$ & $6.5e-1$ \\
    \hline
    \end{tabular}
    \label{tab:scaling}
\end{table*}

\subsubsection{Representative examples}

Next we provide some representative behaviors of the algorithm’s performance in response to different problem setups. 

\begin{figure*}[htbp]
    \centering
    \begin{subfigure}[b]{0.45\textwidth}
        \centering
        \includegraphics[width=\textwidth]{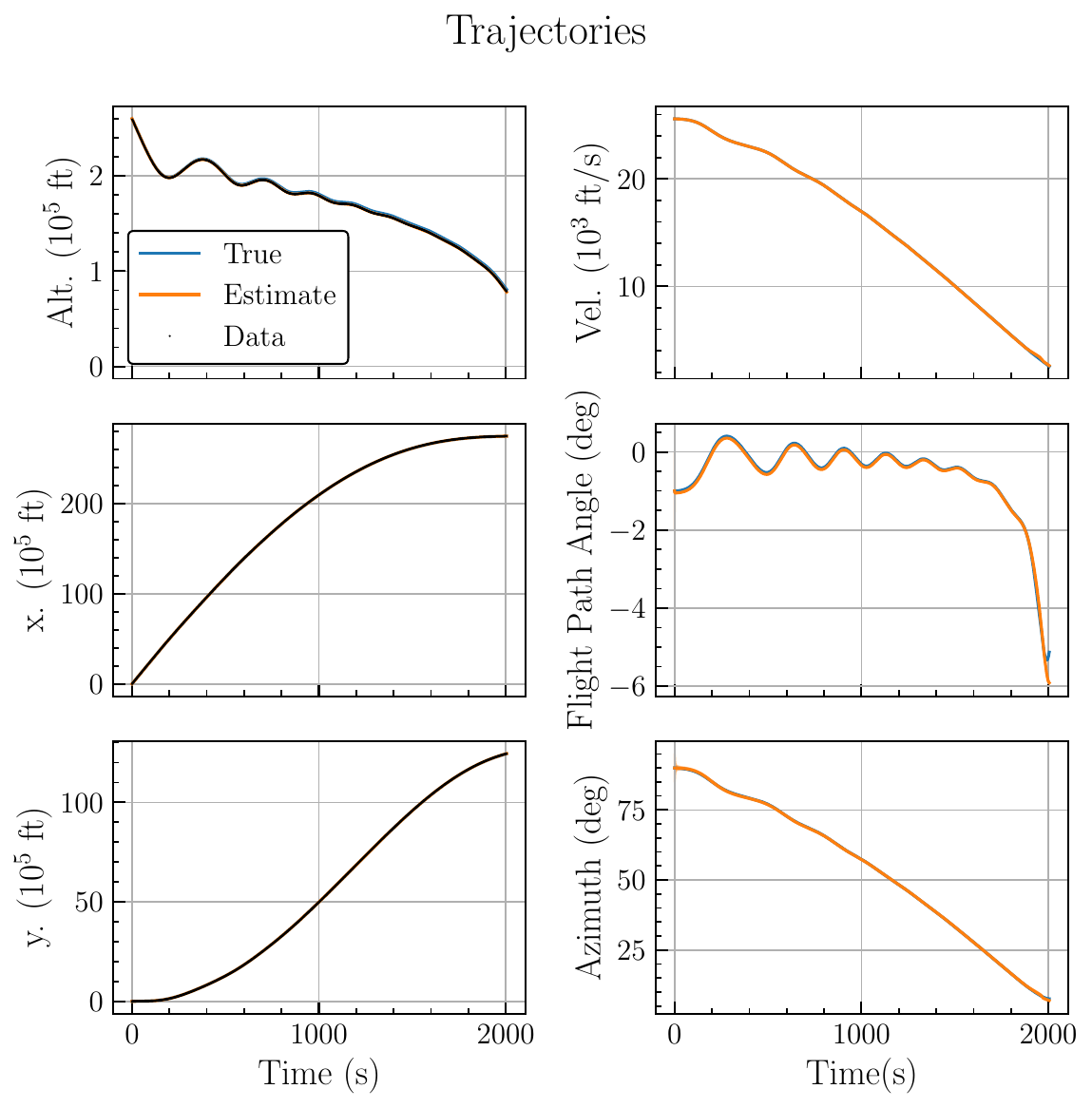}
        \caption{Estimated States}
    \end{subfigure}
    \hfill
    \begin{subfigure}[b]{0.45\textwidth}
        \centering
        \includegraphics[width=\textwidth]{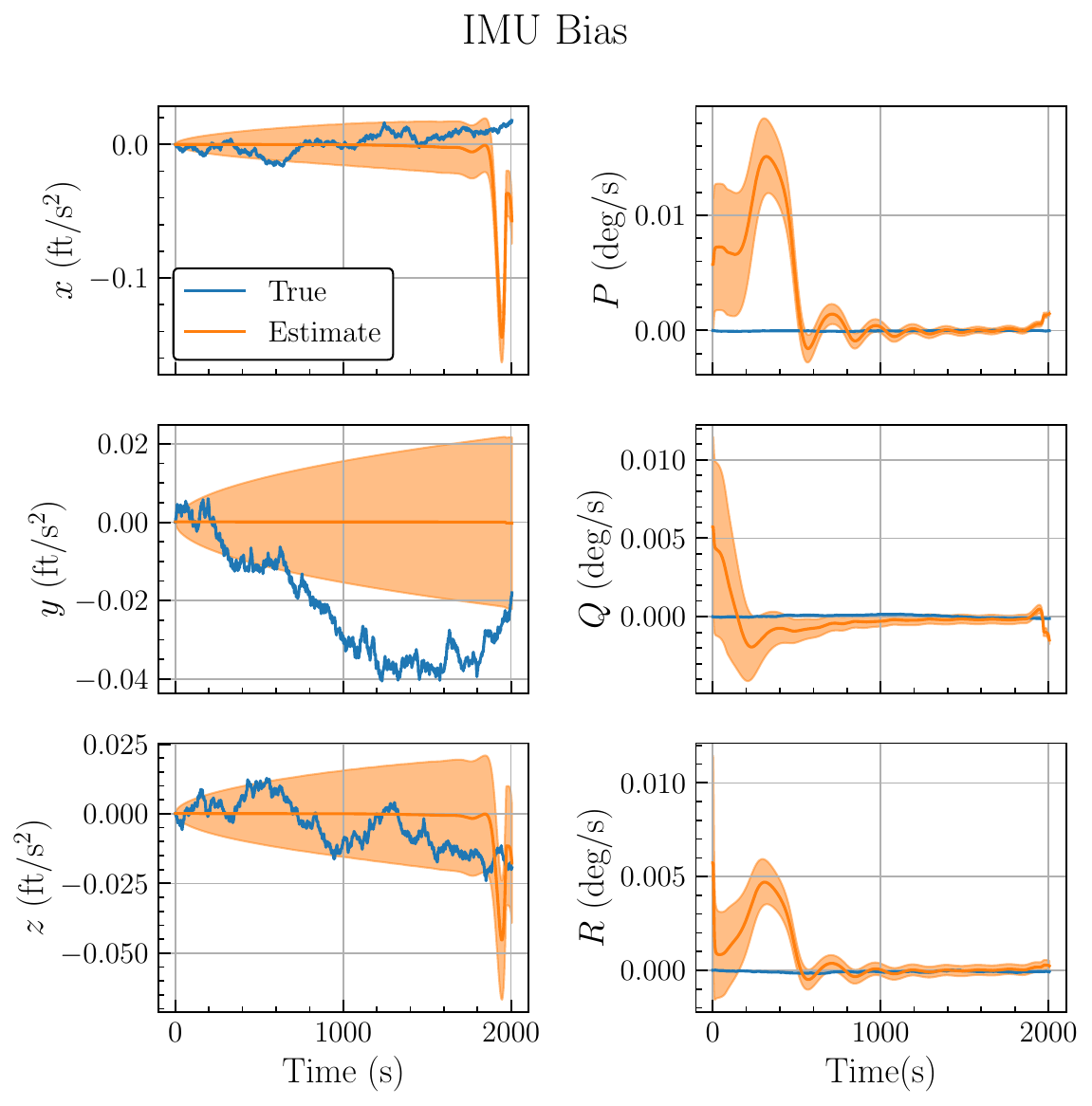}
        \caption{IMU Bias Estimate}
    \end{subfigure}
    \vfill
    \begin{subfigure}[b]{0.45\textwidth}
        \centering
        \includegraphics[width=\textwidth]{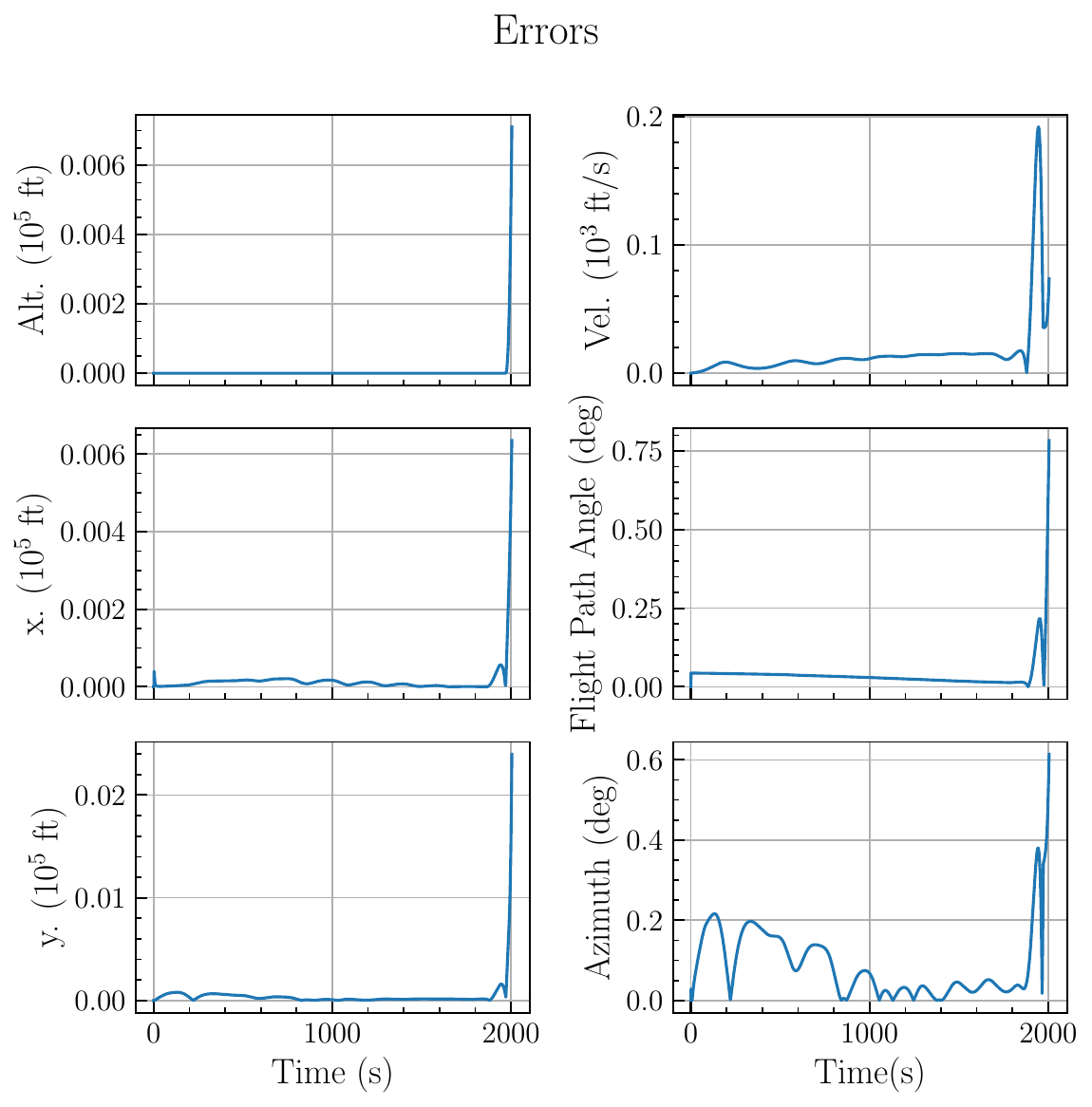}
        \caption{Absolute Errors}
    \end{subfigure}
    \hfill
    \begin{subfigure}[b]{0.45\textwidth}
        \centering
        \includegraphics[width=\textwidth]{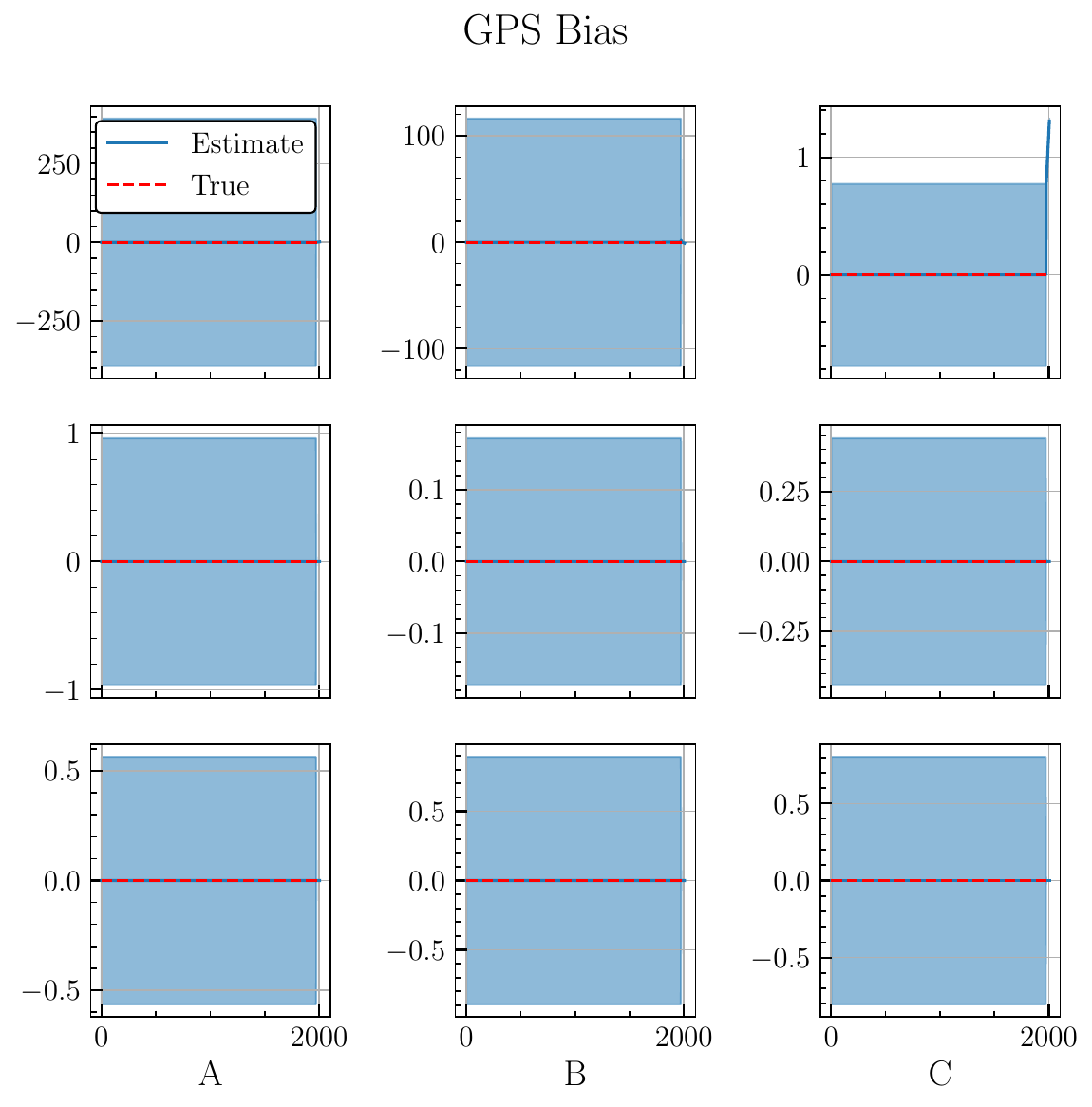}
        \caption{GPS Bias parameters}
    \end{subfigure}
    \caption{Shuttle Reentry Test 1: Unbiased Simulation. The GPS bias plot demonstrates static (A), linear (B), and quadratic (C) corruption parameters for altitude (first row), longitude (second row), and latitude (third row) states of the system. In the absence of corruption, the method does not start learning parameters until the very end of the simulation, and performs traditional Gaussian filtering with functioning sensor readings. One of the corrupted branches' likelihood dominates the likelihood of unbiased branch, resulting in estimated corruption at the end of the simulation. }
    \label{fig:test1}
\end{figure*}

In Test 1 we examine the unbiased case, by setting all three corruption model parameters equal to zero. The purpose of this test is to verify that the filter can still perform with regular, uncorrupted GPS measurements. The results of this test are plotted in the Figure \ref{fig:test1}. As seen in the plots and in the table, the estimated switching time is at the very end of the simulation, meaning that the filter uses the unbiased model throughout most of the time. It is also important to note, that at the last time step, there is no peaked probability distribution among all final branches. The branches that survive until the end have switching times of 1954.72, 1956.12, 1970.11, 1975.71, and  2005.09 seconds with corresponding likelihoods of 50066, 50121, 50425, 50381, 47269. In other words, all surviving branches other than the unbiased one start corruptions at the very end of the simulation and have nearly identical likelihoods. As discussed before, only the unbiased branch has no corruption, and every other branch does identify some sort of bias. Consequently, since all survived branches use the corruption starting at the very end, it indicates that there was no corruption present, and that SKF identifies the unbiased case correctly. 

\begin{figure*}[htbp]
    \centering
    \begin{subfigure}[b]{0.45\textwidth}
        \centering
        \includegraphics[width=\textwidth]{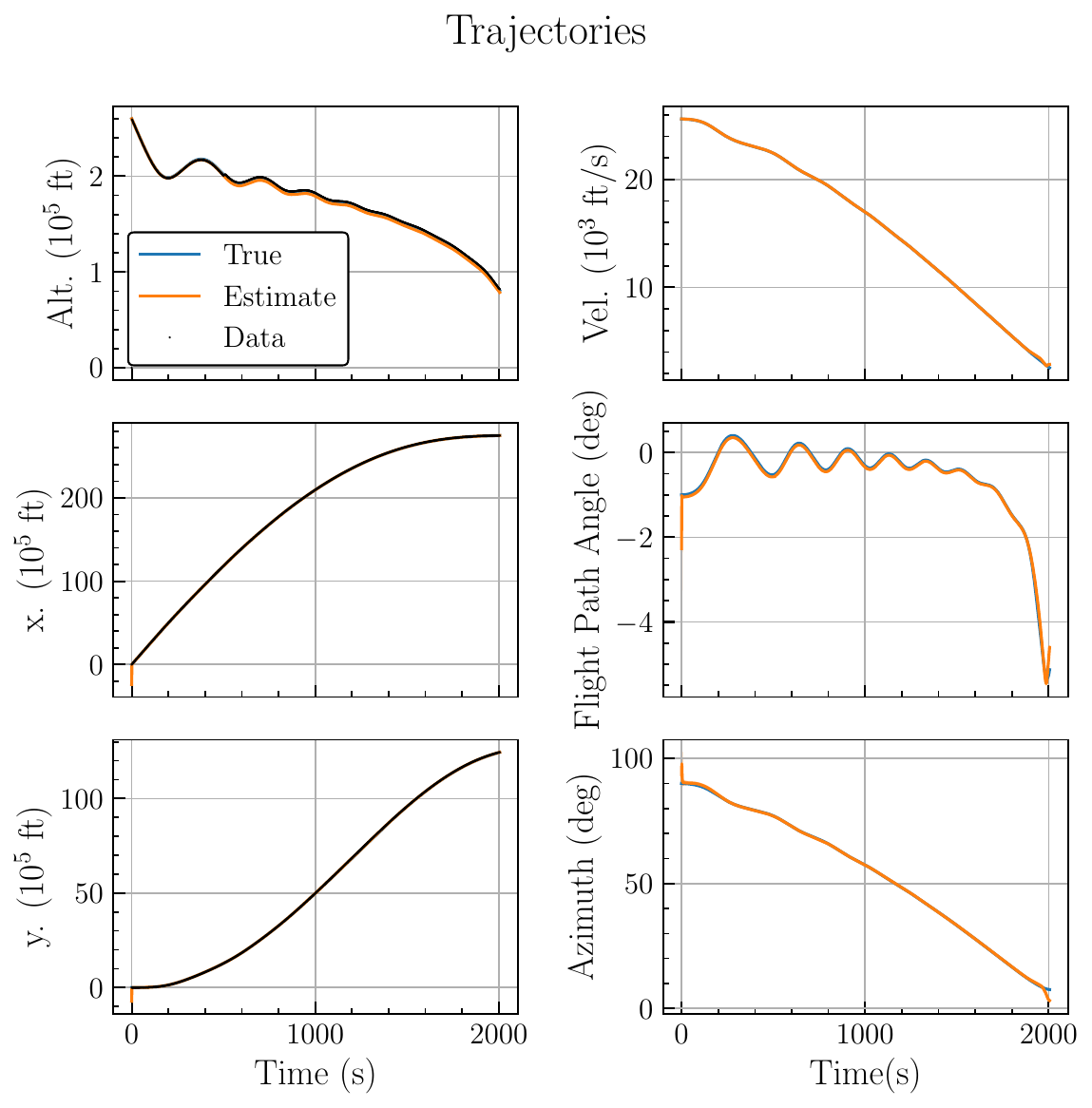}
        \caption{Estimated States}
    \end{subfigure}
    \hfill
    \begin{subfigure}[b]{0.45\textwidth}
        \centering
        \includegraphics[width=\textwidth]{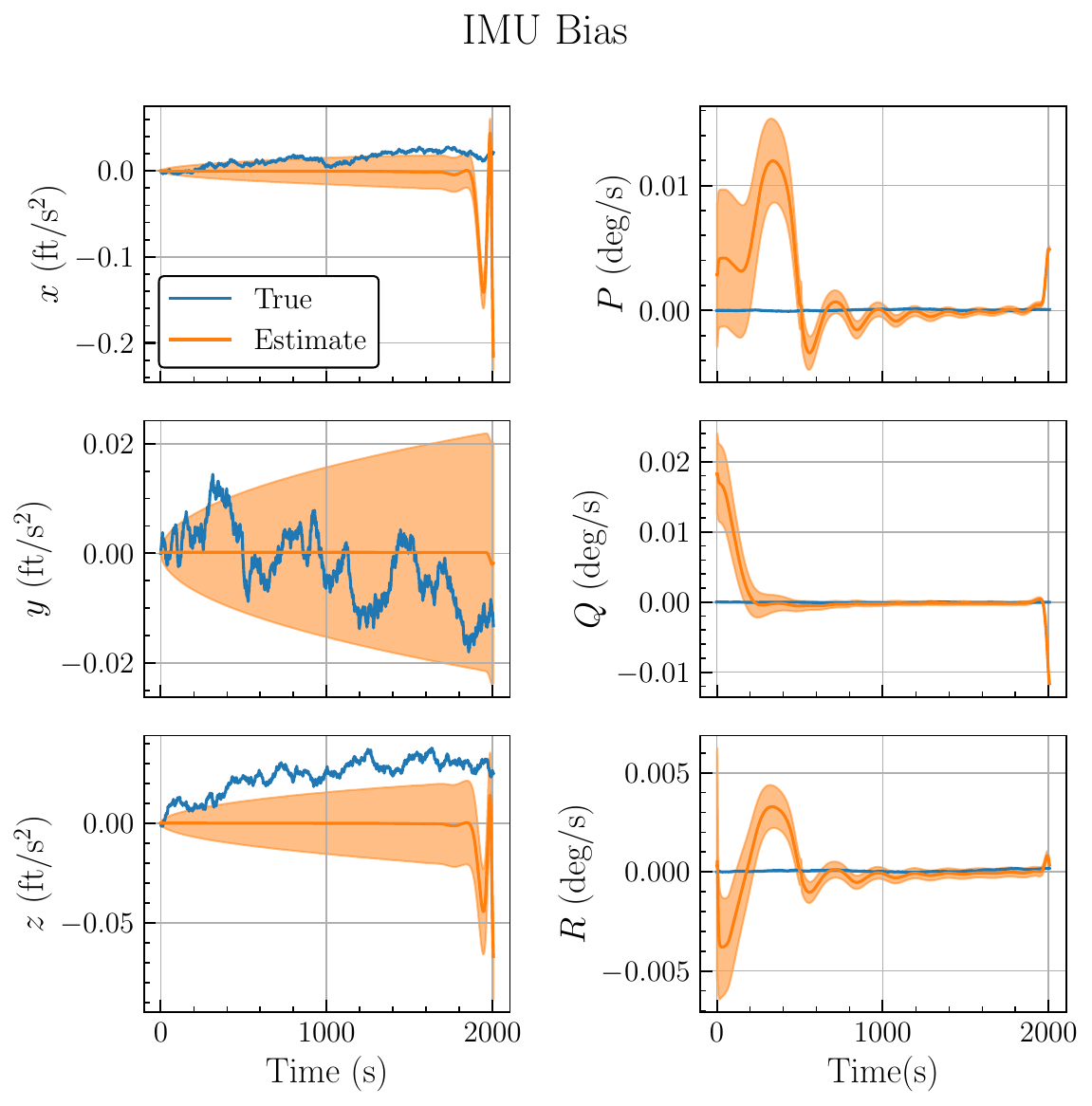}
        \caption{IMU Bias Estimate}
    \end{subfigure}
    \vfill
    \begin{subfigure}[b]{0.45\textwidth}
        \centering
        \includegraphics[width=\textwidth]{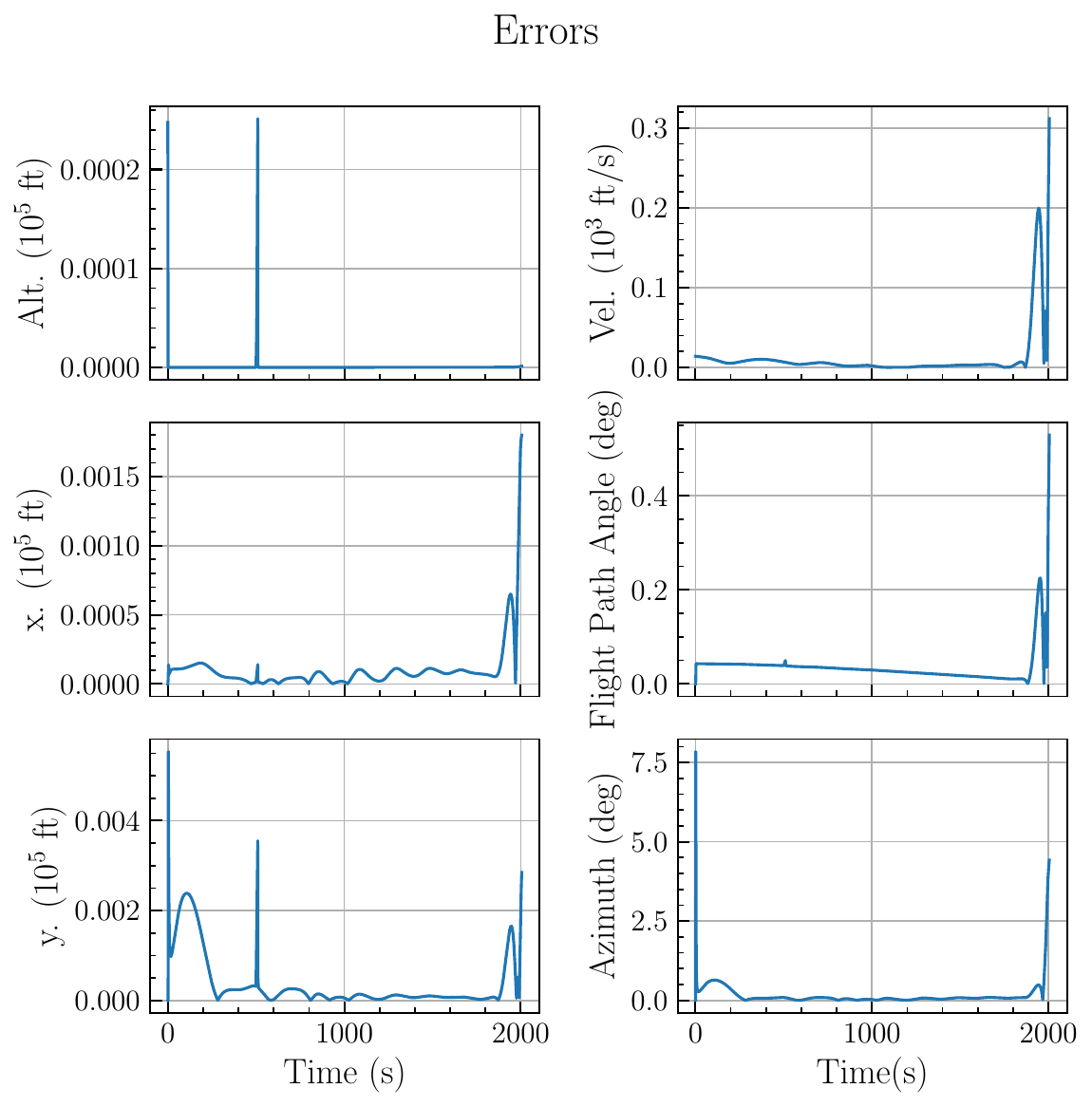}
        \caption{Absolute Errors}
    \end{subfigure}
    \hfill
    \begin{subfigure}[b]{0.45\textwidth}
        \centering
        \includegraphics[width=\textwidth]{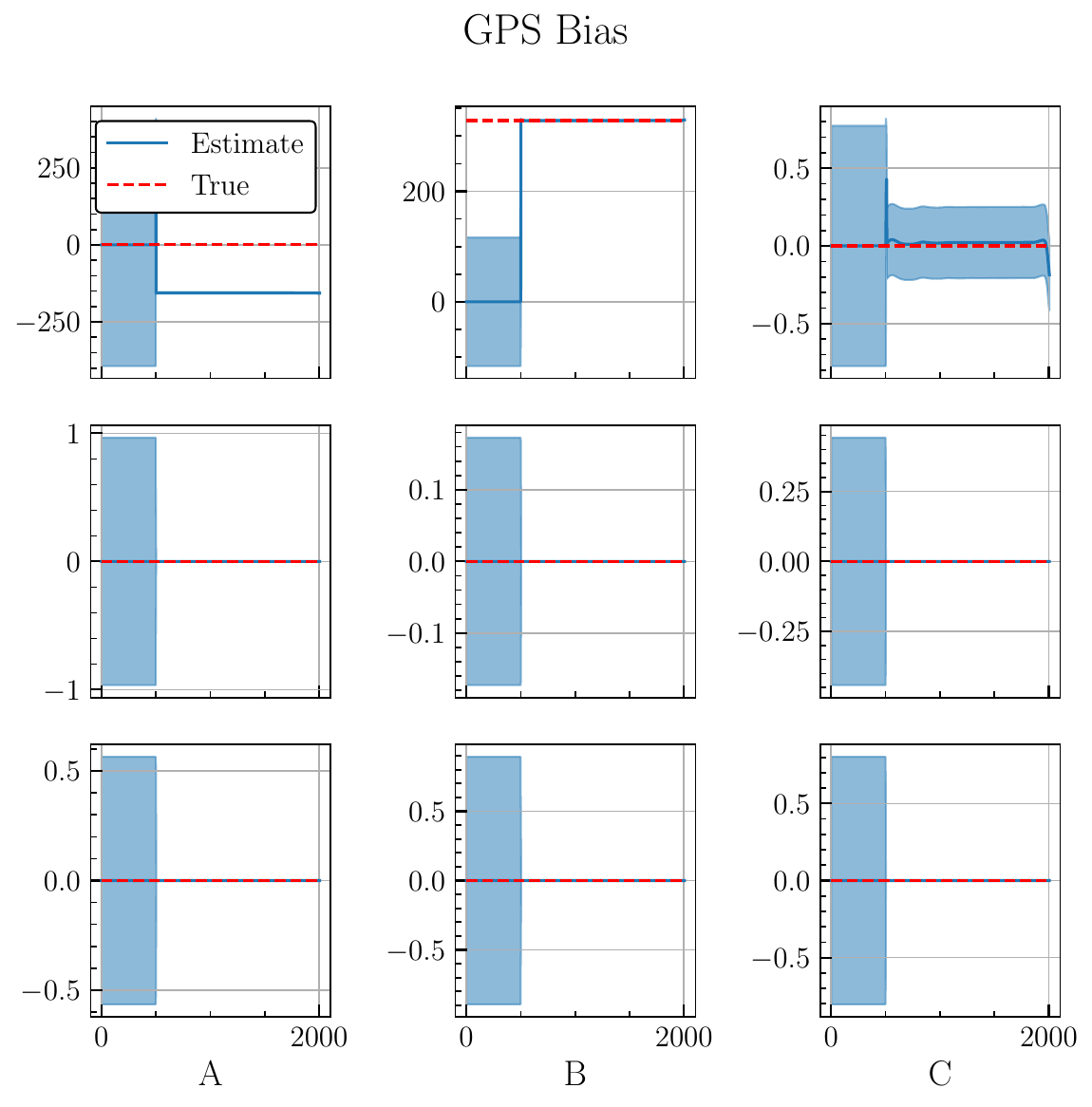}
        \caption{GPS Bias parameters}
    \end{subfigure}
    \caption{Shuttle Reentry Test 10: Simulation with large bias parameters. The GPS bias plot demonstrates static (A), linear (B), and quadratic (C) corruption parameters for altitude (first row), longitude (second row), and latitude (third row) states of the system. With large bias parameters the method identifies the corruption start almost exactly. Knowing the switching time it can accurately estimate the initial state.}
    \label{fig:test10}
\end{figure*}

Next we examine the effect on parameter magnitude on the performance of the proposed algorithm. As seen in Table \ref{tab:shuttle_results}, when applied to the data generated with smaller corruption parameters (Tests 2-9) the SKF does not identify the switching time correctly for all the tests. This occurs because the magnitude of the corruption grows very slowly, which results in either slightly delayed estimated switching time (Tests 2,3,5,6, and 9) or the inability to identify a presence of a switch altogether (Tests 4 and 7). However as seen in Tests 10-24 (excluding Test 19) as the bias parameters increase, the algorithm demonstrates great performance by identifying the switching time within half a time step. The plots obtained from Test 10 are demonstrated in Figure \ref{fig:test10}.

Building on these findings, in Tests 25 - 30 we examine the effect of the measurement noise on the performance of SKF. To aid with this investigation we rerun cases 1, 6, and 10 (unbiased, biased that is not captured, and biased that is captured) to quantify the impact of increasing or decreasing the measurement noise. However, after running these tests, for all three cases, we see very little effect in both the estimated switching times and the errors of the simulations remain similar between increased, decreased, and original noises. 

Finally, we conduct Tests 31 - 36 to quantify the effect of process noise on the performance of SKF. In this series of tests we see a clear improvement in ability to capture the switching time correlated with decrease in noise levels for the hard to capture case (Tests 6, 32, and 35). With larger noise, the filter failed to detect the presence of corruption and found the last time step to be the most likely case. As for the decreased noise, the filter almost perfectly captured the switching time. Hence, these results suggest that the SKF can capture more difficult cases (where the corruption increases slowly ) when the process noise decreases. 

\subsubsection{Statistical analysis}

Similar to the balloon problem, we use this section to describe the results of the statistical analysis to further quantify the effects of problem setup parameters on the ability of SKF to accurately estimate the switching time and reference shuttle reentry trajectory. We conduct 1280 experiments varying parameter process noise, measurement noise, and the corruption parameters. The range of values assigned to each of these variables, is listed in Table \ref{tab:test_values_shuttle_sa}. All combinations of these variables are used in the test set and the state process noise is a fixed function of the measurement noise

\begin{equation}
    q_x = \frac{r}{100}
\end{equation}

\begin{table}[ht!]
    \centering
    \caption{Values of test setup variables for the statistical analysis of the shuttle reentry problem}
    \begin{tabular}{|c|c|}
    \hline
    Variable & Tested Values \\
    \hline
    Parameter process noise ($q_p$) & [1e-14, 1e-12, 1e-11, 1e-10] \\
    Measurement Noise ($r$) & [1e-6, 1e-7, 1e-8, 1e-9, 1e-10] \\
    Static Bias Parameter ($A$) & [0, 10, 100, 1000] \\
    Linear Bias Parameter ($B$)& [0, 1, 10, 100] \\
    Quadratic Bias Parameter ($C$)& [0, 0.1, 1, 10] \\
    \hline
    \end{tabular}
    \label{tab:test_values_shuttle_sa}
\end{table}

In the following we summarize the performance of the SKF corresponding to each of these variables in figure pairs: the first figure representing the percentage of successful simulations, where all but the chosen variable is varied; and the second figure of a pair plots the median RMSE obtained over all these other simulations.

We summarize the performance of the SKF corresponding to the magnitudes of each variable using two figures. The first figure representing the percentage of successful simulations, where all but the chosen variable is varied; The second figure is a three by three subplot containing the medium $RMSE$ for each of the states in Eq. (\ref{eq:state}). 

\begin{figure}[ht!]
    \centering
        \includegraphics[width=0.49\textwidth]{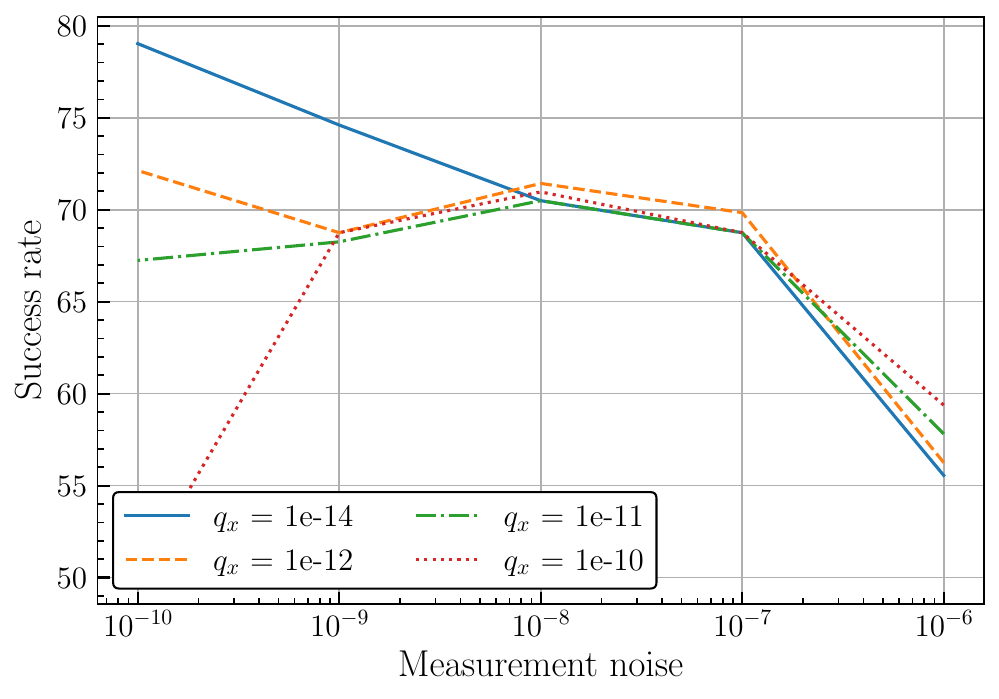}
        \caption{Successful simulations as a function of measurement noise
        for fixed values of parameter process noise. Decreased process noise corresponds to an increase success rate of identifying the switching time. Lower measurement noises tend to result in higher success rates.}
    \label{fig:sa_shut_meas_success}
\end{figure}

Figure \ref{fig:sa_shut_meas_success} demonstrates the success rate of SKF in capturing true switching time as a function of measurement noise. We see that in a region from $10^-8$ to $10^-6$, the success rate behaves similar to that of a balloon example: a decrease in measurement noise results in an increased success rate. However, once we reach that value, depending on the parameter process noise, the success rate either stays around that value for small process noises( $q_x = 1e-14, 1e-12$) or decreases for higher process noises ($q_x = 1e-11, 1e-10$). 

\begin{figure}[ht!]
    \centering
        \includegraphics[width=0.49\textwidth]{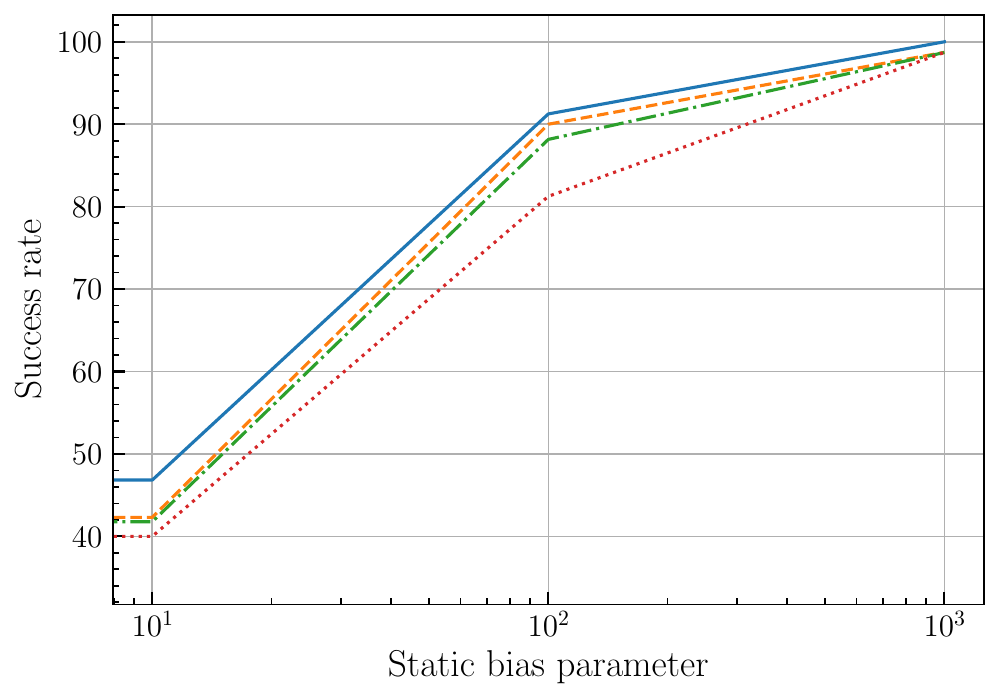}
        \caption{Successful simulations as a function of static corruption parameter
        for fixed values of parameter process noise. Larger static parameters correspond to an increased
        success rate of identifying switching time.}
    \label{fig:sa_shut_stat_success}
\end{figure}
\begin{figure}[ht!]
    \centering
        \includegraphics[width=0.49\textwidth]{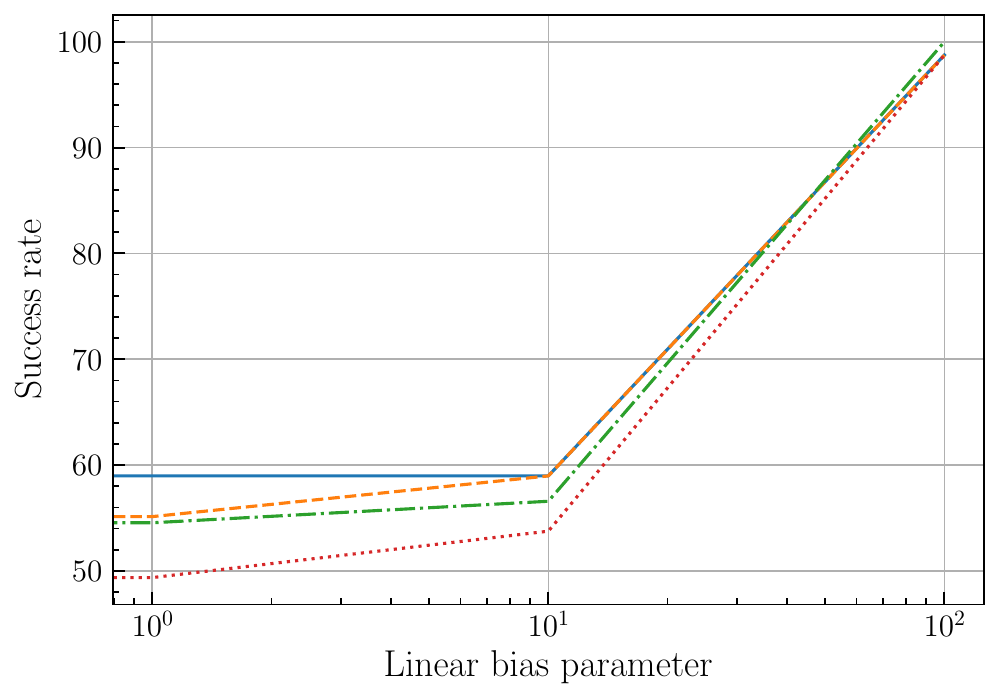}
        \caption{Successful simulations as a function of linear corruption parameter
        for fixed values of parameter process noise. Larger linear parameters correspond to an increased
        success rate of identifying switching time.}
    \label{fig:sa_shut_lin_success}
\end{figure}
\begin{figure}[ht!]
    \centering
        \includegraphics[width=0.49\textwidth]{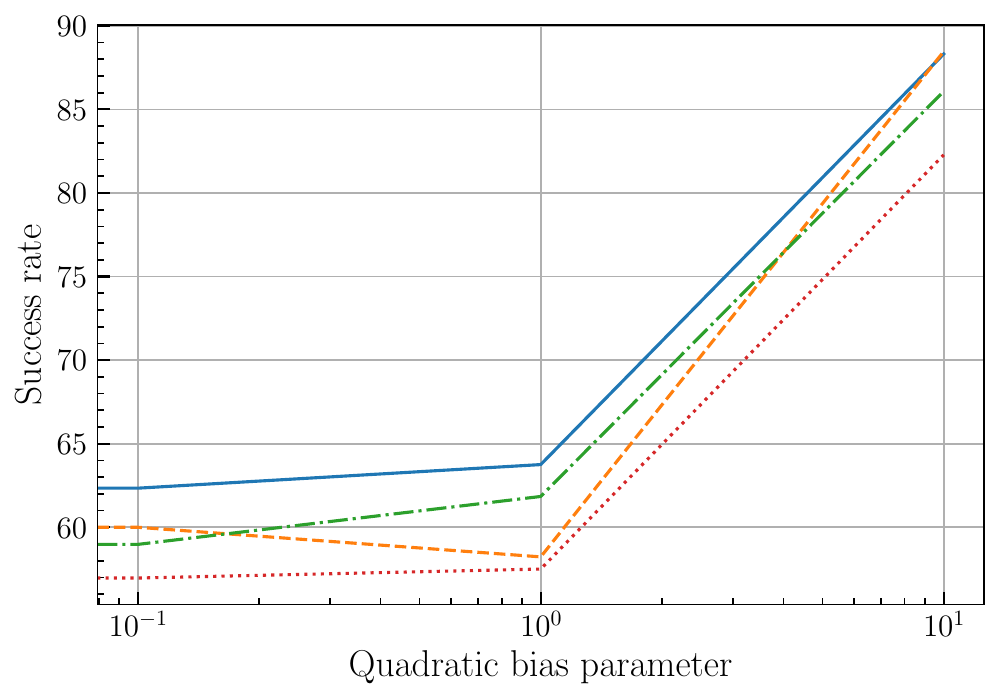}
        \caption{Successful simulations as a function of quadratic corruption parameter
        for fixed values of parameter process noise. Larger quadratic parameters correspond to an increased
        success rate of identifying switching time.}
    \label{fig:sa_shut_quad_success}
\end{figure}

Figures \ref{fig:sa_shut_stat_success}, \ref{fig:sa_shut_lin_success} , \ref{fig:sa_shut_quad_success} illustrate the success rate of SKF in capturing true switching time as a function of corruption parameters. Similar to the balloon example, an increase in bias parameters demonstrate an improvement in method's ability to capture the time at which the corruption begins. This behavior is present regardless of process noise magnitude. 

\begin{figure*}[ht!]
    \centering
        \includegraphics[width=\textwidth]{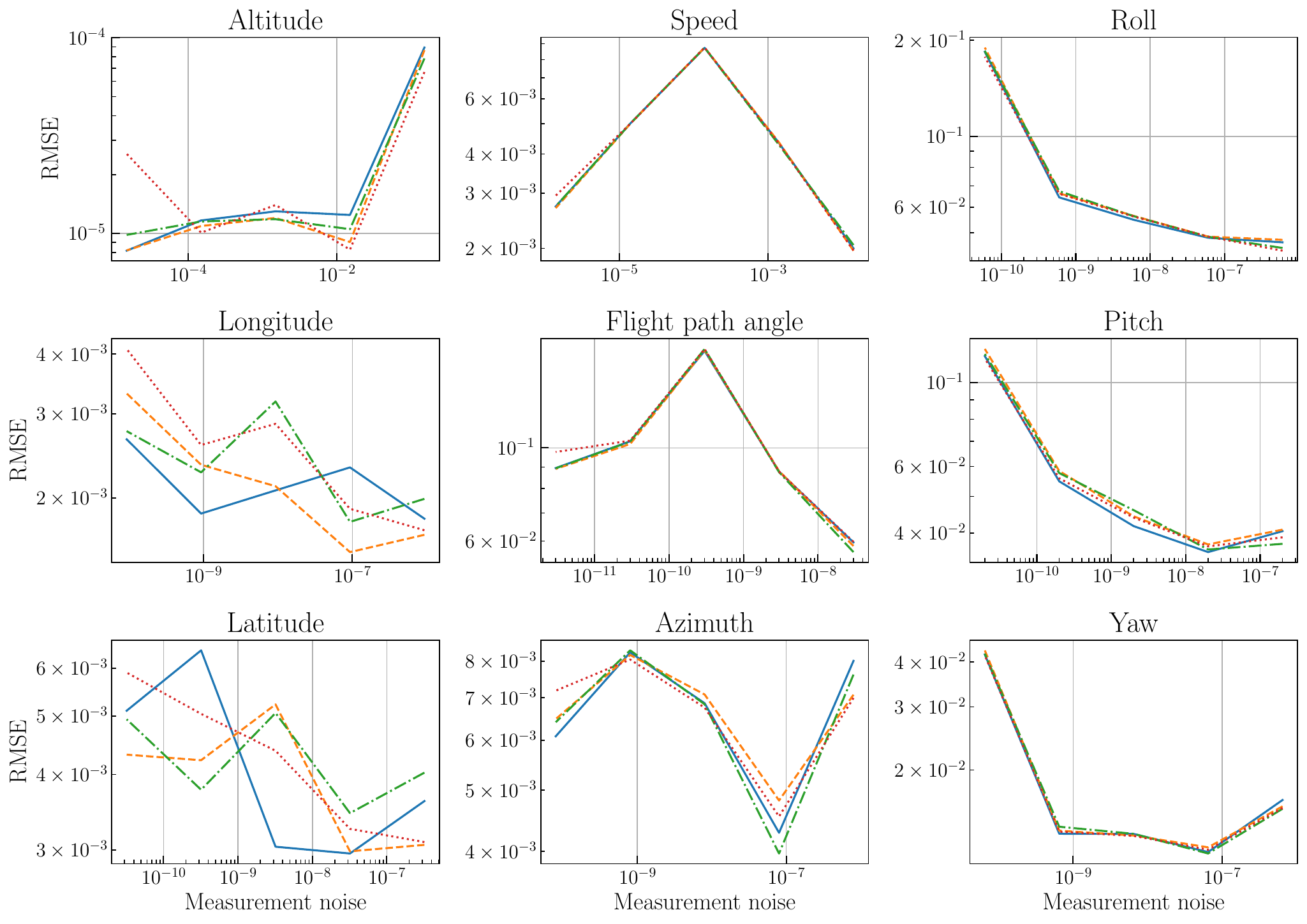}
        \caption{$RMSE$ between estimated states and inertial model for each of the states as a function of measurement noise for fixed values of parameter process noise. The $RMSE$ remains fairly constant over these noise levels.}
    \label{fig:sa_shut_meas}
\end{figure*}

Figure \ref{fig:sa_shut_meas} shows how measurement noise affects the $RMSE$ for each of the states in shuttle state vector. The plots do not demonstrate a common pattern that would describe the effect of measurement noise on all of the parameters. 

\begin{figure*}[ht!]
    \centering
        \includegraphics[width=\textwidth]{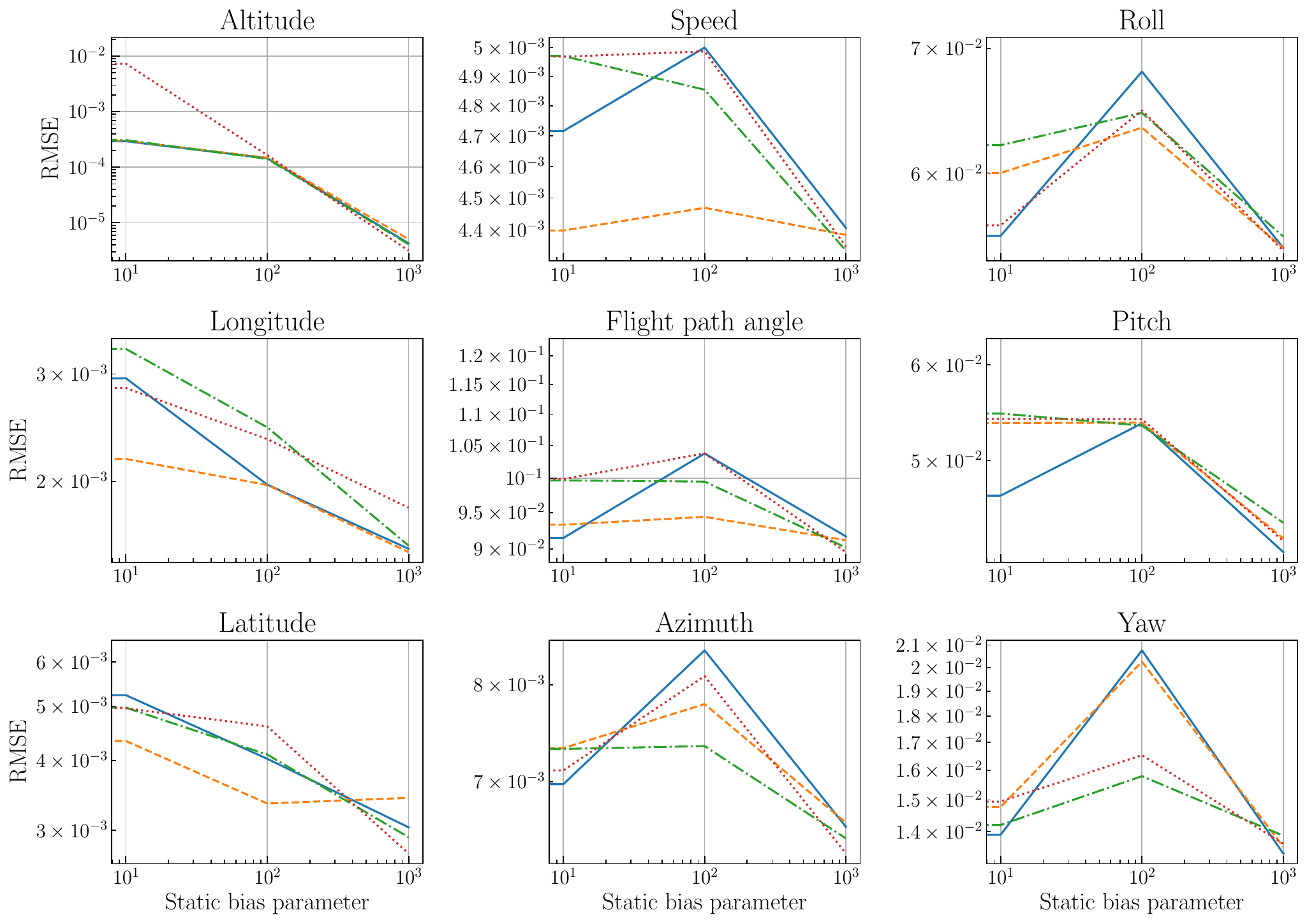}
        \caption{$RMSE$ between estimated states and inertial model for each of the states as a function of static corruption parameter for fixed values of parameter process noise. For most states, there is a peaked value in $RMSE$ corresponding to a state where corruption is significant enough to introduce errors but not significant enough to identify the switching time.}
    \label{fig:sa_shut_static}
\end{figure*}
\begin{figure*}[ht!]
    \centering
        \includegraphics[width=\textwidth]{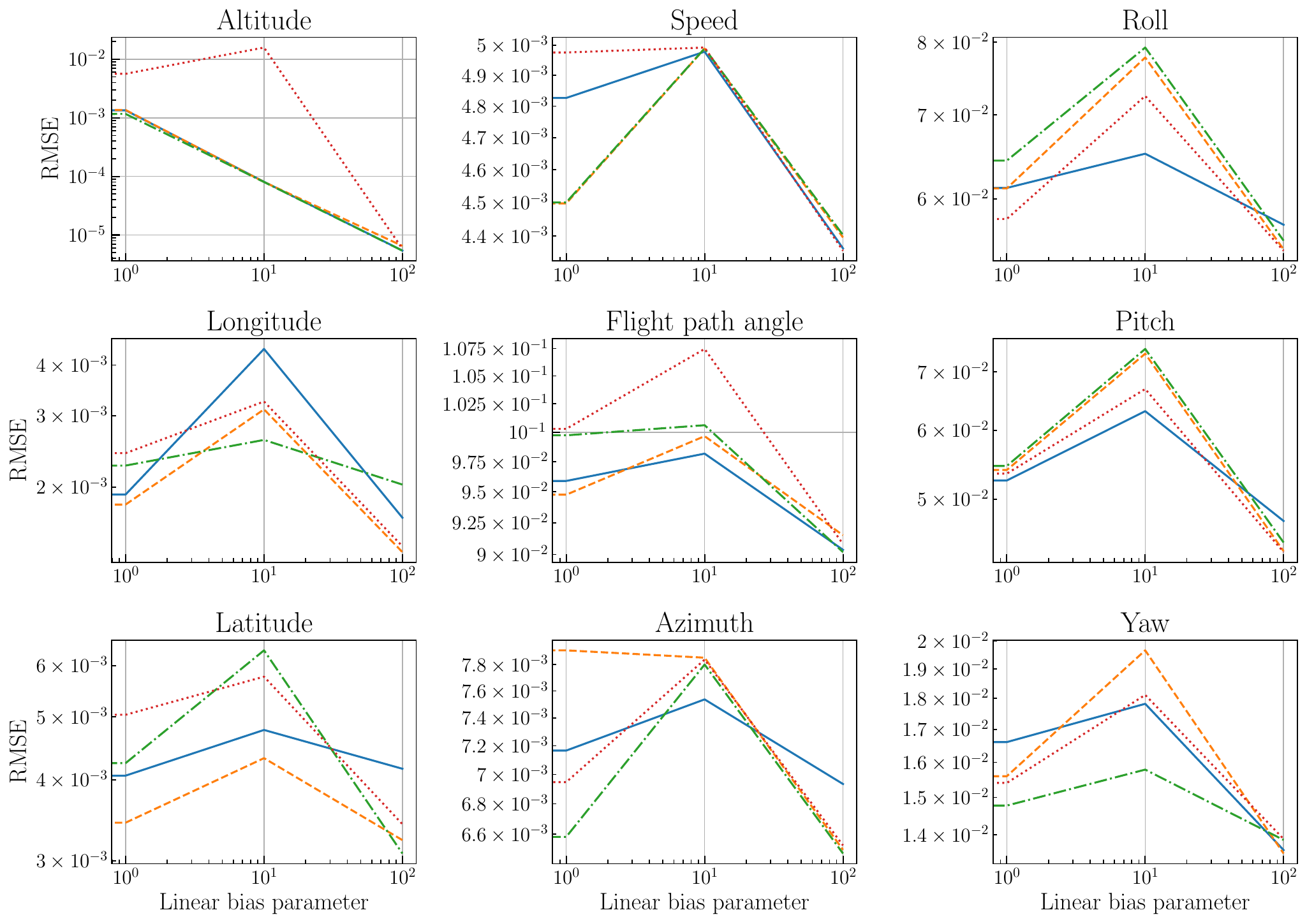}
        \caption{$RMSE$ between estimated states and inertial model for each of the states as a function of linear corruption parameter for fixed values of parameter process noise. For most states, there is a peaked value in $RMSE$ corresponding to a state where corruption is significant enough to introduce errors but not significant enough to identify the switching time.}
    \label{fig:sa_shut_lin}
\end{figure*}
\begin{figure*}[ht!]
    \centering
        \includegraphics[width=\textwidth]{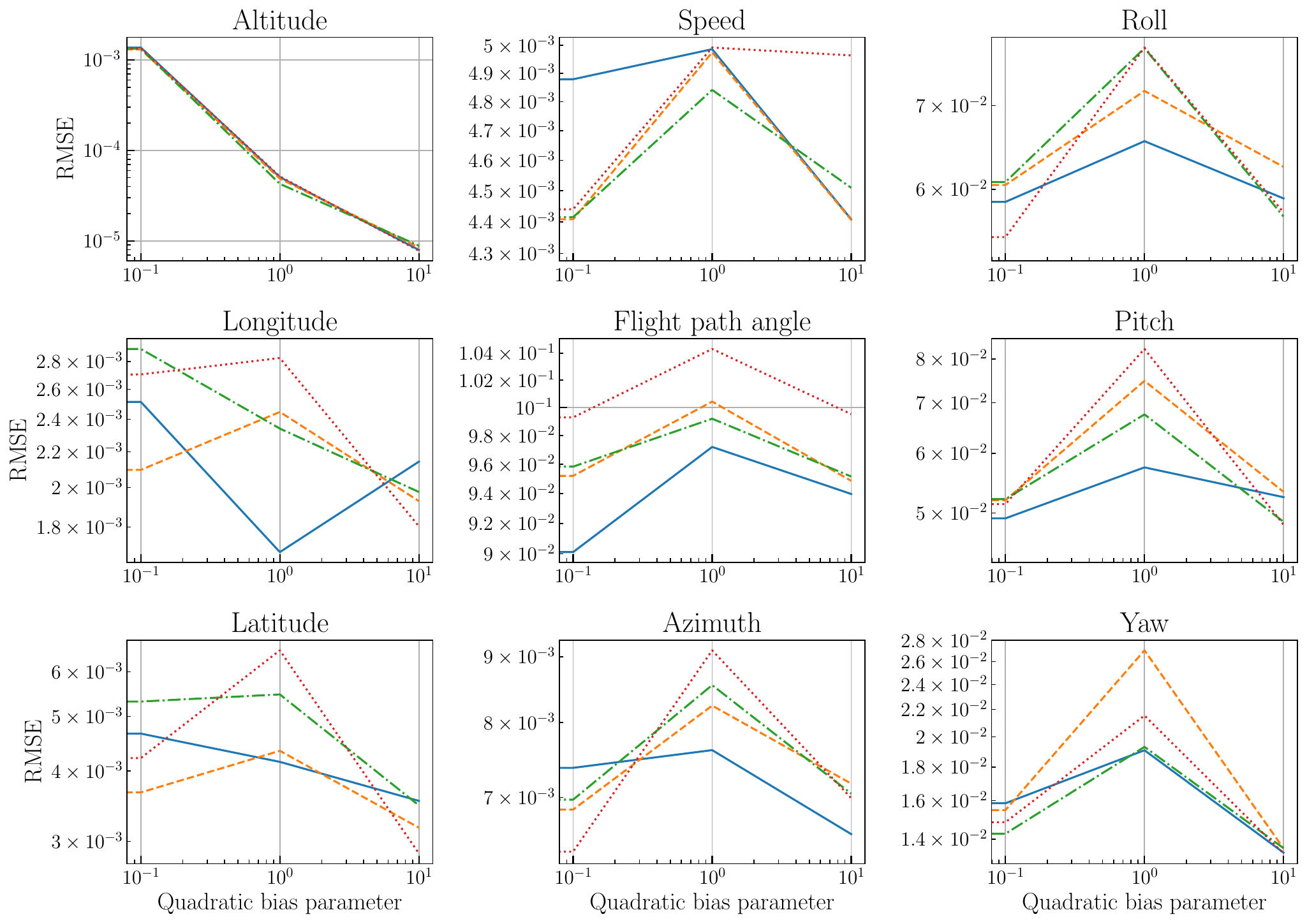}
        \caption{$RMSE$ between estimated states and inertial model for each of the states as a function of quadratic corruption parameter for fixed values of parameter process noise. For most states, there is a peaked value in $RMSE$ corresponding to a state where corruption is significant enough to introduce errors but not significant enough to identify the switching time.}
    \label{fig:sa_shut_quad}
\end{figure*}

Figures \ref{fig:sa_shut_static}, \ref{fig:sa_shut_lin}, \ref{fig:sa_shut_quad} illustrate the impact of corruption parameters on the RMSE of the shuttle state vector relative to the true shuttle trajectory. The majority of the sub-figures exhibit a similar trend: errors are low for both extreme ends of the corruption parameter range, while peaking in the middle. As previously discussed in the balloon example, this occurs because, with small corruption parameters, the bias is insufficient to significantly mislead the estimation, resulting in consistently low errors, even when the corruption is not detected. However, as the bias increases, it is more readily identified by the state augmentation approach, which facilitates easier correction.
 
Overall, we see that in cases where the bias is large, we are able to reliably identify the bias and correct for its errors. When the bias is small, the state estimation error
is also small. It is possible that there is a middle regime where the bias is just large enough to cause an error, but to remain undetected and uncorrected.

\section{Conclusion}\label{sec:conclusion}

In this paper, we introduce a novel approach to state estimation that can accurately estimate state using corrupted measurements. Our approach that combines Switching Kalman Filter (SKF) with parameter augmentation techniques was tested on two critical applications associated with balloon navigation and shuttle reentry problems. Our numerical results demonstrate that the proposed approach significantly improves the robustness and accuracy of state estimation in the presence of GPS data corruption,  however, the accuracy obtained depends on several factors. For example, the proposed method demonstrated significantly better performance in identifying biases with large corruption parameters and in cases where the corruption was initiated later in the simulation. Additionally, lower state and parameter process noises, along with higher sampling frequencies, improved the method’s ability to accurately capture the true switching time. Overall, the proposed methodology enhances the reliability of state estimation systems by retaining and processing potentially corrupted data rather than discarding it, leading to more accurate and resilient state estimates. Future work will focus on further refining the algorithm to reduce computational costs and exploring its application to other domains where sensor data integrity is critical.

\section*{Acknowledgements}

We would like to thank Tucker Haydon for providing his implementation for the IMU model used in this paper and Connor Brashar for his guidance choosing relevant test cases for the algorithms proposed in this paper.
This work was funded in part by Sandia National Laboratories’ Laboratory Directed Research and Development (LDRD) program and by an NSF CAREER Award CMMI-2238913.
This article has been co-authored by employees of National Technology and Engineering Solutions of Sandia, LLC under Contract No. DE-NA0003525 with the U.S. Department of Energy (DOE). The employees co-own right, title and interest in and to the article and are responsible for its contents. The United States Government retains and the publisher, by accepting the article for publication, acknowledges that the United States Government retains a non-exclusive, paid-up, irrevocable, world-wide license to publish or reproduce the published form of this article or allow others to do so, for United States Government purposes. The DOE will provide public access to these results of federally sponsored research in accordance with the DOE Public Access Plan \url{https://www.energy.gov/downloads/doe-public-access-plan}.

\bibliographystyle{plain}
\bibliography{references}
\end{document}